\documentclass[longauth]{aa}
\usepackage{subcaption}
\usepackage{graphicx}
\usepackage{xcolor}

\usepackage{txfonts}
\usepackage{hyperref}
\usepackage{xspace}
\hypersetup{
    colorlinks=true,
    linkcolor=blue,
    citecolor=blue,
    filecolor=magenta,      
    urlcolor=cyan,
    pdftitle={OASIS II: The Ionized Gas Traces Feedback-Driven Declines into the Quiet Side of Bursty Star Formation at z > 4},
    pdfpagemode=FullScreen,
    }
\usepackage{lastpage}
\makeatletter
\renewcommand*\aa@pageof{, page \thepage{} of \pageref*{LastPage}}
\makeatother

\newcommand{\nii}{[N\,\textsc{ii}]}
\newcommand{\sii}{[S\,\textsc{ii}]}

\newcommand{\oii}{[O\,\textsc{ii}]}
\newcommand{\oiii}{[O\,\textsc{iii}]}

\newcommand{\neiii}{[Ne\,\textsc{iii}]}

\newcommand{\civ}{C\,\textsc{iv}}

\newcommand{\hb}{H\,$\beta$}

\newcommand{\Mstar}{\ensuremath{M_\star}\xspace}

\let\oldAA\AA
\renewcommand{\AA}{\oldAA\xspace}

\newcommand{\jwst}{\textit{JWST}\xspace}

\newcommand{\ppxf}{{\sc ppxf}\xspace}

\begin{document}

   \title{OASIS II: The Ionized Gas Traces Feedback-Driven Declines into the Quiet Side of Bursty Star Formation at z > 4}

    \titlerunning{OASIS II: Feedback-driven declines and the ionized gas}

\author{
Tobias J.\ Looser\inst{1}\fnmsep\thanks{tobias.looser@cfa.harvard.edu}
\and Francesco D'Eugenio\inst{2,3}
\and William M.\ Baker\inst{4}
\and Santiago Arribas\inst{5}
\and Andrew J.\ Bunker\inst{6}
\and Stefano Carniani\inst{7}
\and St\'ephane Charlot\inst{8}
\and Jacopo Chevallard\inst{6}
\and Mirko Curti\inst{9}
\and Qiao Duan\inst{2,3}
\and Kevin Hainline\inst{10}
\and Jakob M.\ Helton\inst{11}
\and Yuki Isobe\inst{2,3,12}
\and Benjamin D.\ Johnson\inst{1}
\and Pierluigi Rinaldi\inst{13,14}
\and Brant Robertson\inst{15}
\and Fengwu Sun\inst{1}
\and Sandro Tacchella\inst{2,3}
\and James A.\ A.\ Trussler\inst{1}
\and Hannah \"Ubler\inst{16}
\and Joris Witstok\inst{17,18}
\and Zihao Wu\inst{1}
}

\institute{
Center for Astrophysics $|$ Harvard \& Smithsonian, 60 Garden St., Cambridge MA 02138, USA
\and Kavli Institute for Cosmology, University of Cambridge, Madingley Road, Cambridge CB3 0HA, UK
\and Cavendish Laboratory, University of Cambridge, 19 JJ Thomson Avenue, Cambridge CB3 0HE, UK
\and DARK, Niels Bohr Institute, University of Copenhagen, Jagtvej 155A, DK-2200 Copenhagen, Denmark
\and Centro de Astrobiolog\'{\i}a (CAB), CSIC--INTA, Cra.\ de Ajalvir Km.~4, 28850 Torrej\'on de Ardoz, Madrid, Spain
\and Department of Physics, University of Oxford, Denys Wilkinson Building, Keble Road, Oxford OX1 3RH, UK
\and Scuola Normale Superiore, Piazza dei Cavalieri 7, I-56126 Pisa, Italy
\and Sorbonne Universit\'e, CNRS, UMR 7095, Institut d'Astrophysique de Paris, 98 bis bd Arago, 75014 Paris, France
\and INAF -- Osservatorio di Astrofisica e Scienza dello Spazio, Via P.\ Gobetti 93/3, I-40129 Bologna, Italy
\and Steward Observatory, University of Arizona, 933 N.\ Cherry Avenue, Tucson, AZ 85721, USA
\and Department of Astronomy \& Astrophysics, The Pennsylvania State University, University Park, PA 16802, USA
\and Waseda Research Institute for Science and Engineering, Faculty of Science and Engineering, Waseda University, 3-4-1 Okubo, Shinjuku, Tokyo 169-8555, Japan
\and Department of Astronomy, The University of Texas at Austin, Austin, TX 78712, USA
\and Cosmic Frontier Center, The University of Texas at Austin, Austin, TX 78712, USA
\and Department of Astronomy and Astrophysics, University of California, Santa Cruz, 1156 High Street, Santa Cruz, CA 95064, USA
\and Max-Planck-Institut f\"ur extraterrestrische Physik (MPE), Gie{\ss}enbachstra{\ss}e 1, 85748 Garching, Germany
\and Cosmic Dawn Center (DAWN), Copenhagen, Denmark
\and Niels Bohr Institute, University of Copenhagen, Jagtvej 128, DK-2200 Copenhagen, Denmark
}

\authorrunning{T.\ J.\ Looser et al.}
   \date{}

\abstract
{Models of bursty star formation in low-mass ($10^{8}$--$10^{9.5}$\,M$_\odot$) galaxies at $z>4$ differ most in the suppressed phases: what drives the decline, how fast star formation falls, and how long it stays low. OASIS Paper~I showed that such quiet galaxies are common and old; here we measure how they got there and how their ionized gas responds. We compare Balmer-line star-formation rates (SFRs) with 100-Myr averages from the stellar continuum, in NIRSpec/PRISM spectra of 54 OASIS galaxies at $z>4$, including three mini-quenched galaxies below the quenching threshold and five lulling galaxies at least 0.6\,dex below the star-forming main sequence (SFMS) but still star-forming. Galaxies above the SFMS rise and fade equally often, whereas those below it are mostly fading: the lulling galaxies formed little more than a tenth of their mass in the last 100\,Myr, and their decline began tens to ${\sim}100$\,Myr before observation. Suppressed phases thus last of order 100\,Myr, as long as or longer than in most simulations, whereas recovery is much faster: few galaxies below the SFMS are rising. The ionized gas tracks these changes in star formation. [O\,\textsc{iii}]/[O\,\textsc{ii}] falls from galaxies above the SFMS to the lulling ones, even at fixed mass and after dust correction, and follows the recent change in SFR ($\mathrm{SFR_{10}}/\mathrm{SFR_{100}}$) more closely than the position relative to the SFMS: two rising galaxies below the SFMS are highly ionized. This is expected if fewer and softer ionizing photons than in strongly star-forming galaxies meet gas depleted by earlier star formation. Lulling galaxies show ISM conditions distinct from those of strongly star-forming galaxies. The declines are faster than passive gas exhaustion predicts, favoring feedback that expels the gas or keeps it from forming stars. Quiet phases are thus long, feedback-driven episodes, and the ionized gas traces where a galaxy is in its burst cycle.}


   \maketitle\nolinenumbers

\section{Introduction}\label{sec:intro}

JWST has shown that star formation in low-mass galaxies at $z>4$ is bursty: the scatter about the star-forming main sequence (SFMS) is large, particularly when the star-formation rate (SFR) is measured on short timescales \citep[e.g.,][see also \citealt{caplar+tacchella2019,Wan+2025}]{Cole+2025,Clarke+2025,Simmonds+2025}, star-formation histories (SFHs) reconstructed from spectra fluctuate \citep[e.g.,][]{Tacchella+2023,Looser+2025}, and galaxies have been caught with their star formation suppressed or temporarily halted \citep[e.g.,][]{Looser+2024Nat,Strait2023,Baker+2025_zap,Trussler+2025,Covelo-Paz+2025,Khullar+2026}. Cosmological simulations predict such cycles as a generic outcome of stellar feedback in shallow potential wells \citep[e.g.,][]{Faucher-Giguere2018,Tacchella+2020,Gelli+2023,McClymont+2025,Katz+2025}, but they differ in the masses and redshifts at which this happens, in how often, and especially in how quickly star formation falls, how long it stays low, and which processes drive the decline \citep{Dome2023,Gelli+2025}. In several simulations, suppressed phases last only a few tens of Myr \citep{Dome2023,Faisst+2024}, whereas longer ones appear in others only when defined less strictly or at lower masses \citep{McClymont+2025,Attar+d2026}.

In the first paper of this series \citep[][hereafter OASIS Paper~I]{Looser+2026_OASISI}, we presented the JWST Cycle~3 program OASIS (GO~5997), which obtained deep NIRSpec/PRISM spectroscopy of galaxies across the full burst cycle at $z>4$, including the suppressed phases that earlier, emission-line-selected samples missed. Of the 54 galaxies in the $z>4$ analysis sample, three are mini-quenched, with specific SFRs below a quenching threshold of $0.2/t_{\rm obs}$, and five lulling, at least 0.6\,dex below the SFMS but above the quenching threshold. Mini-quenched galaxies typically show no detectable Balmer emission, and lulling galaxies weak but measurable emission. These suppressed galaxies are common: they make up ${\sim}20$ per cent of the galaxies at $\log_{10}(M_\star/M_\odot)=8$--9.5, and 29 per cent at $z\leq5.5$ in the target-priority tiers without selection on star-formation state, corresponding to a number density of ${\sim}1.6\times10^{-3}$\,Mpc$^{-3}$\,dex$^{-1}$ in an order-of-magnitude estimate. More robustly, they are old: their Balmer breaks and stellar ages rise with distance below the SFMS, and the mini-quenched galaxies have light-weighted ages above 100\,Myr. They show little stellar reddening, and they lie at masses where most simulations produce few suppressed galaxies.

OASIS Paper~I thus established how common and how old the suppressed galaxies are, but not how they became suppressed: how fast their star formation declined, when the decline began, and which mechanism removed or neutralized the gas that fuelled it. These are the most direct tests of feedback models, but they are hard to answer from a single snapshot, because the instantaneous SFR alone does not reveal whether a galaxy is forming stars at a steady rate or whether its star formation is rising or fading, and photometric SED fits recover the recent change only loosely (OASIS Paper~I).

Several mechanisms could drive the decline. Low-mass halos at these redshifts are fed by cold streams at inflow rates per unit halo mass far higher than today (Sect.~\ref{sec:mechanisms}), so that suppression requires the gas to be removed or heated, or its supply interrupted. Stellar feedback can expel gas in outflows with mass-loading factors of order unity to tens \citep{Muratov+2015,Pandya+2021}, although supernova feedback alone may act too slowly to truncate star formation abruptly \citep{Gelli+2023b}; radiation pressure may clear the star-forming regions of the most intensely star-forming galaxies \citep{Ferrara+2024}; and accreting black holes, now found in substantial numbers at $z>4$ \citep[e.g.,][]{Kocevski+2023,Harikane+2023,Maiolino2024,Juodzbalis+2026}, can eject or heat the gas even in dwarf galaxies \citep{koudmani+2022,Arjona-Galvez+2024}. Ionized outflows are common in low-mass galaxies at $z>3$ \citep{Carniani+2024_outflows}, and neutral-gas outflows capable of quenching have been observed in quiescent galaxies \citep[e.g.,][]{Belli+2024,DEugenio+2024NatAs...8.1443D,Wu+2025,Valentino+2025,Oh+2026,Sapori+2026,Zhu+2026}, but so far only in massive ones. Distinguishing between these mechanisms in low-mass galaxies requires knowing how fast their star formation declines and how the properties of their gas change as it does.

A direct way to measure how star formation is changing is to compare tracers sensitive to different timescales: nebular emission traces the last ${\sim}10$\,Myr, whereas the UV continuum integrates over a longer but less well-defined timescale, and the rest-frame optical stellar continuum over ${\sim}100$\,Myr and more \citep{FloresVelazquez+2021}. Diagnostics of this kind have been applied to simulations
\citep[e.g.,][]{Sparre+2017}, to photometric samples
\citep[e.g.,][]{Ciesla+2024, Atek+2022, Asada+2024, Navarro-Carrera+2024,
Endsley+2025, Carvajal-Bohorquez+2025}, and to spectroscopic ones
\citep[e.g.,][]{Broussard+2019, Endsley+2023a, Clarke+2025, Looser+2025, Perry+2025, Ulivi+2026}. These samples are, however, dominated by star-forming galaxies, so that the suppressed side of the cycle, where the nebular lines are weakest, remains poorly sampled.

The interstellar medium offers a second handle. Spectroscopy at $z>4$ has so far focused on UV-bright, emission-line-dominated galaxies. These galaxies exhibit the hallmarks of extreme starbursts: high gas densities \citep[e.g.,][]{reddy+2023}, elevated ionization parameters traced by extreme $\mathrm{O32} \equiv [\mathrm{O\,\textsc{iii}}]\lambda5007/[\mathrm{O\,\textsc{ii}}]\lambda\lambda3726,3729$ and $\mathrm{Ne3O2} \equiv [\mathrm{Ne\,\textsc{iii}}]\lambda3869/[\mathrm{O\,\textsc{ii}}]\lambda\lambda3726,3729$ ratios \citep[e.g.,][]{Nakajima+2023,Cameron+2023,Sanders+2023,Trump+2023}, high $[\mathrm{O\,\textsc{iii}}]\lambda5007/\mathrm{H}\beta$ excitation \citep[e.g.,][]{Curti+2024,Nakajima+2023,Cameron+2023,Trump+2023,Sanders+2023}, and very low metallicities \citep[e.g.,][]{Curti+2024,Nakajima+2023,Trump+2023,Brinchmann+2023}. These conditions match theoretical expectations for compact, metal-poor star-forming regions, which naturally generate ionization parameters far above those of $z\sim0$ galaxies \citep{Kewley+2013,Gutkin+2016,Nakajima+2014,Plat+2019}. 

In star-forming galaxies, the ionization parameter rises with the specific SFR at fixed mass \citep{Kaasinen+2018,Papovich+2022,Shen+2025}, out to $z\approx9$ \citep{Cleri+2026}, and in simulations the ISM follows the burst cycle, with lower H\,\textsc{ii}-region densities after a burst \citep{Katz+2023} and O32 tracing the recent SFH \citep{Choustikov+2023}. Observationally, in NIRSpec/PRISM spectra of star-forming galaxies at $z=3.5$--7, those whose star formation has recently declined, selected by a low H$\alpha$-to-UV ratio, lie below the [O\,\textsc{iii}]/H$\beta$ expected from the mass--metallicity relation, whereas O32 shows no correlation with this indicator \citep{Cooper+2026}. Whether the ionization follows the current position relative to the SFMS or its recent change in SFR, and what happens when star formation declines, has not yet been tested with an indicator that traces the recent change robustly, nor in suppressed galaxies below the SFMS, whose lines are weak and which have been largely absent from such samples.

In this paper we measure directly from the spectra whether, and how fast, star formation is rising or fading across the full burst cycle, including in the suppressed galaxies: we compare the Balmer-line SFR with the SFR averaged over the last 100\,Myr from full spectral fitting of the stellar continuum. This places every galaxy on a track through the SFR--$M_\star$ plane and dates the onset of its suppression. We then test whether the ionization of the gas follows the position of a galaxy relative to the SFMS or its recent change in star-formation activity, and we combine both results to constrain the mechanisms of suppression. Specifically, we address the following key questions:
\begin{itemize}
    \item How fast does star formation decline in suppressed galaxies at $z>4$, and how long ago did the decline begin?
    \item Does the state of the ionized gas follow the recent change in star formation?
    \item Can passive gas consumption explain the decline, or is feedback required, and of what kind?
\end{itemize}

Section~\ref{sec:Methods} summarizes the data and methods, which are described in detail in OASIS Paper~I; a third companion paper \citep[][hereafter OASIS Paper~III]{Baker+2026_OASISIII} relates burstiness to galaxy size. Section~\ref{sec:decline} presents the decline of star formation and the duration and frequency of suppressed phases, Sect.~\ref{sec:bpt} the emission-line diagnostics, and Sect.~\ref{sec:mechanisms} discusses what suppresses star formation. Section~\ref{sec:Summary} summarizes our conclusions. Throughout, we adopt a \citet{chabrier_galactic_2003} initial mass function and a flat $\Lambda$CDM cosmology with $H_0=69.6$\,km\,s$^{-1}$\,Mpc$^{-1}$, $\Omega_{\rm M}=0.286$ and $\Omega_\Lambda=0.714$.

\section{Data and Methods}\label{sec:Methods}

The data, sample and measurements are described in detail in OASIS Paper~I; we summarize here what is needed for this paper.

\subsection{Observations and sample}\label{sec:sample}

OASIS (GO~5997) observed two pointings in GOODS-S \citep{giavalisco2004}, in the Hubble Ultra Deep Field \citep{Beckwith+2006} and the JADES Origins Field \citep{Eisenstein2023}, with the NIRSpec \citep{jakobsen+2022} micro-shutter array \citep[MSA;][]{Ferruit2022} and the PRISM, which covers 0.6--5.3\,$\mu$m at a resolution $R\approx30$--330. The MSA configurations were designed with \texttt{eMPT} \citep{Bonaventura+2023}, and each target received 9.3--28\,h of exposure. The spectra were reduced with the pipelines of the ESA NIRSpec Science Operations Team and the NIRSpec GTO Team \citep{Carniani+2024_outflows,Scholtz2025_JADES_DR4}. Of the 65 galaxies with a secure redshift $z\gtrsim4$, 54 have the robust estimates of stellar mass, stellar-population properties and nebular SFR that the analysis requires; they form the $z>4$ sample used throughout.

We classify the galaxies by their distance from the star-forming main sequence, $\Delta_{\rm MS}=\log\mathrm{SFR_{neb}}-\log\mathrm{SFR_{MS}}(M_\star,z)$, with the nebular SFR of Sect.~\ref{sec:sfr}, measured against the SFMS of \citet{Simmonds+2025} at each galaxy's own redshift, and by the quenching threshold $\mathrm{sSFR_{neb}}<0.2/t_{\rm obs}(z)$, where $t_{\rm obs}$ is the cosmic time at the redshift of observation. Mini-quenched galaxies lie below this threshold, lulling galaxies above it but at $\Delta_{\rm MS}\leq-0.6$, and nearly-MS galaxies at $-0.6<\Delta_{\rm MS}\leq0$; we refer to the remainder as galaxies above the SFMS. The sample contains three mini-quenched, five lulling, 17 nearly-MS galaxies and 29 above the SFMS (Fig.~\ref{fig:SFR-mass}). The mini-quenched and lulling galaxies, together the suppressed galaxies, make up 15 per cent of the sample, 21 per cent at $\log_{10}(M_\star/M_\odot)=8$--9.5, and 19 per cent in the priority tiers that carry no information on star-formation state (OASIS Paper~I). Spectra and physical properties of all mini-quenched and lulling galaxies are presented in OASIS Paper~I; their emission-line fluxes and line ratios are listed in Table~\ref{tab:lines}.

\subsection{Star-formation rates, stellar masses and SFHs}\label{sec:sfr}

The nebular SFR, $\mathrm{SFR_{neb}}$, which traces star formation over the last ${\sim}10$\,Myr, is derived from the dust-corrected H$\alpha$ luminosity with the calibration of \citet{shapley2023}. Dust is corrected with the Balmer decrement and the SMC bar extinction curve \citep{Gordon+2003} where H$\alpha$ and H$\beta$ are both detected, and otherwise with the stellar reddening from the continuum fit, scaled by a nebular-to-stellar ratio of 1.3 \citep[cf.][]{Song+2026,Karthikeyan+2026}. At $z>7.1$, where H$\alpha$ falls outside the PRISM range, H$\beta$ is converted to H$\alpha$ with the Case~B ratio of 2.86 and corrected with the scaled stellar reddening. Galaxies without significant Balmer lines are assigned $3\sigma$ upper limits.

Stellar masses and SFHs come from full spectral fitting with \ppxf\ \citep{Cappellari2017,cappellari2022}, which models the continuum as a non-negative combination of simple stellar populations (SSPs), built from C3K atmospheres \citep{Conroy2019} and MIST isochrones \citep{Choi2016} and attenuated with the \citet{Calzetti2000} law, and the emission lines as Gaussians. Masses are converted from a Salpeter to a Chabrier IMF \citep{speagle+2014}, and uncertainties are derived from bootstrap resampling \citep{Looser+2024_FMR,Looser+2025}. The recovered SSP weights are robust when summed over 1-dex bins in age, but the weights of adjacent SSPs are not individually reliable (OASIS Paper~I). Beyond the mean
stellar ages, we use only quantities that integrate the SFH:
the mass formed within the last 100\,Myr, which gives the SFR averaged over that window, $\mathrm{SFR_{ppxf}}(100\,\mathrm{Myr})$ (the median over the bootstrap realizations, used throughout), and the cumulative
fraction of mass formed within a lookback time $T$, $f(T)$. As a cumulative sum, $f(T)$ is affected only by trade-offs between the SSPs adjacent to $T$, which are included in its bootstrap uncertainties (Fig.~\ref{fig:mass_fraction}). Unlike the UV-continuum tracers used in most previous studies (Sect.~\ref{sec:intro}), the full spectral fit draws simultaneously on all the information in the spectrum, including the UV brightness, the slope $\beta$, the strength of the Balmer break and the dust attenuation.

We define the recent change in star formation as $\Delta\log\mathrm{SFR}\equiv\log(\mathrm{SFR_{neb}}/\mathrm{SFR_{ppxf}}(100\,\mathrm{Myr}))$: positive for galaxies whose star formation has risen relative to the last 100\,Myr and negative for those that have faded (Fig.~\ref{fig:SFR-mass}). Where $\mathrm{SFR_{neb}}$ is an upper limit, so is $\Delta\log\mathrm{SFR}$. For galaxies with an upper limit on $\mathrm{SFR_{neb}}$ and a very low $\mathrm{SFR_{ppxf}}(100\,\mathrm{Myr})$, such as the low-mass mini-quenched galaxy 206569 (Fig.~\ref{fig:SFR-mass}; OASIS Paper~I), $\Delta\log\mathrm{SFR}$ is therefore only weakly constrained: its upper limit can be positive even if star formation has declined.

Because $\Delta\log\mathrm{SFR}$ is set by two largely independent measurements, from the nebular Balmer lines and directly from the stellar continuum, it is free of the uncertainties that affect ratios of model age bins.  The two rates are on a common scale: $\mathrm{SFR_{neb}}$ agrees with the independent photometric Prospector rate over 10\,Myr \citep{Simmonds+2025} to a median $-0.05$\,dex (normalized median absolute deviation, NMAD, 0.16\,dex) for the galaxies above the SFMS and $-0.09$\,dex (NMAD 0.21\,dex) for all 37 galaxies with both rates, and with $\mathrm{SFR_{ppxf}}(100\,\mathrm{Myr})$ to $-0.11$\,dex in the median over the 49 with a detected nebular SFR. $\mathrm{SFR_{ppxf}}(100\,\mathrm{Myr})$ in turn agrees with the Prospector rate over the same 100\,Myr to $+0.13$\,dex (NMAD 0.28\,dex, 41 galaxies). The scatter between the two timescales is larger (NMAD 0.55\,dex for the 49) and largely physical: for the 37 galaxies with all four rates, the two matched-timescale comparisons scatter by only 0.21 and 0.27\,dex, which together account for at most 0.34\,dex of the 0.51\,dex scatter in their $\Delta\log\mathrm{SFR}$, and Prospector's own ratio of its 10 and 100\,Myr rates correlates with $\Delta\log\mathrm{SFR}$ (Spearman $\rho=+0.57$), although its continuity prior damps such changes \citep{leja+2019a}.

The stellar continuum alone shows the same changes, at least qualitatively: the ratio of the \ppxf\ SFRs over the last 10 and the preceding 90\,Myr correlates with $\Delta\log\mathrm{SFR}$ ($\rho=+0.56$, $p=2\times10^{-5}$, 42 galaxies, against ${\approx}0.13$ expected from their shared continuum term) and is lower below the SFMS than above it ($p=0.02$), but its level is offset by ${\sim}0.7$\,dex and it is too noisy to classify individual galaxies. The Balmer break follows $\Delta\log\mathrm{SFR}$ more closely ($\rho=-0.61$ for the 36 galaxies with a well-measured break) than $\Delta_{\mathrm{MS}}$ ($\rho=-0.51$, and only $-0.06$ at fixed $\Delta\log\mathrm{SFR}$), and not at all the offset of $\mathrm{SFR_{ppxf}}(100\,\mathrm{Myr})$ from the SFMS ($\rho=+0.03$). The break thus reflects a recent rise or decline rather than the amount of older stars: a rise fills it in with new O and early-B stars, whereas after a decline it emerges within tens of Myr, once these have died and later-B stars dominate the light.

\subsection{Emission lines and statistics}\label{sec:lines}

Line fluxes are those of the Gaussian components of the \ppxf\ fits. We treat a line as detected if its flux exceeds $3\sigma$, where $\sigma$ is the noise integrated over the line profile; otherwise we assign it a $3\sigma$ upper limit. We base our analysis on four ratios formed from \oii, \neiii, \hb\ and \oiii:
\begin{align}
  \mathrm{O32} &= \frac{[\mathrm{O\,III}]\lambda5007}{[\mathrm{O\,II}]\lambda\lambda3726,3729},
  \label{eq:O32}\\
  \mathrm{Ne3O2} &= \frac{[\mathrm{Ne\,III}]\lambda3869}{[\mathrm{O\,II}]\lambda\lambda3726,3729},
  \label{eq:Ne3O2}\\
  \mathrm{R3} &= \frac{[\mathrm{O\,III}]\lambda5007}{\mathrm{H}\beta},
  \label{eq:R3}\\
  \mathrm{R23} &= \frac{[\mathrm{O\,III}]\lambda\lambda4959,5007+[\mathrm{O\,II}]\lambda\lambda3726,3729}{\mathrm{H}\beta},
  \label{eq:R23}
\end{align}
where R23 uses the fitted fluxes of both [O\,\textsc{iii}] lines, whose ratio is fixed at the
theoretical value of 2.98 in the line fits \citep{Storey+2000}. We quote all four ratios as decimal logarithms, formed from observed fluxes; we discuss the effect of dust in Sect.~\ref{sec:bpt}. We do not base our analysis on the classical line-ratio diagrams \citep[e.g., the BPT diagram;][]{baldwin+1981}, which constrain the ionizing source rather than the ionization state and which the PRISM data constrain poorly: at the
resolution of the PRISM, H$\alpha$ and \nii\ are blended, and \sii, [O\,\textsc{i}] and the
auroral [O\,\textsc{iii}]$\lambda4363$ line are detected in only a minority of our galaxies and
in none of the suppressed ones. We show the \sii\ diagram and the \citet{Mazzolari+2024} [O\,\textsc{iii}]$\lambda4363$ diagnostic in Appendix~\ref{app:bpt}; a robust excitation analysis requires deep grating spectroscopy (Sect.~\ref{sec:ion_source}). We quantify trends with the Spearman rank correlation coefficient $\rho$ and, at fixed stellar mass, with partial rank correlations, as in OASIS Paper~I.

\section{Results: The decline of star formation below the main sequence}\label{sec:decline}

\begin{figure}
    \centering
    \includegraphics[width=\linewidth]{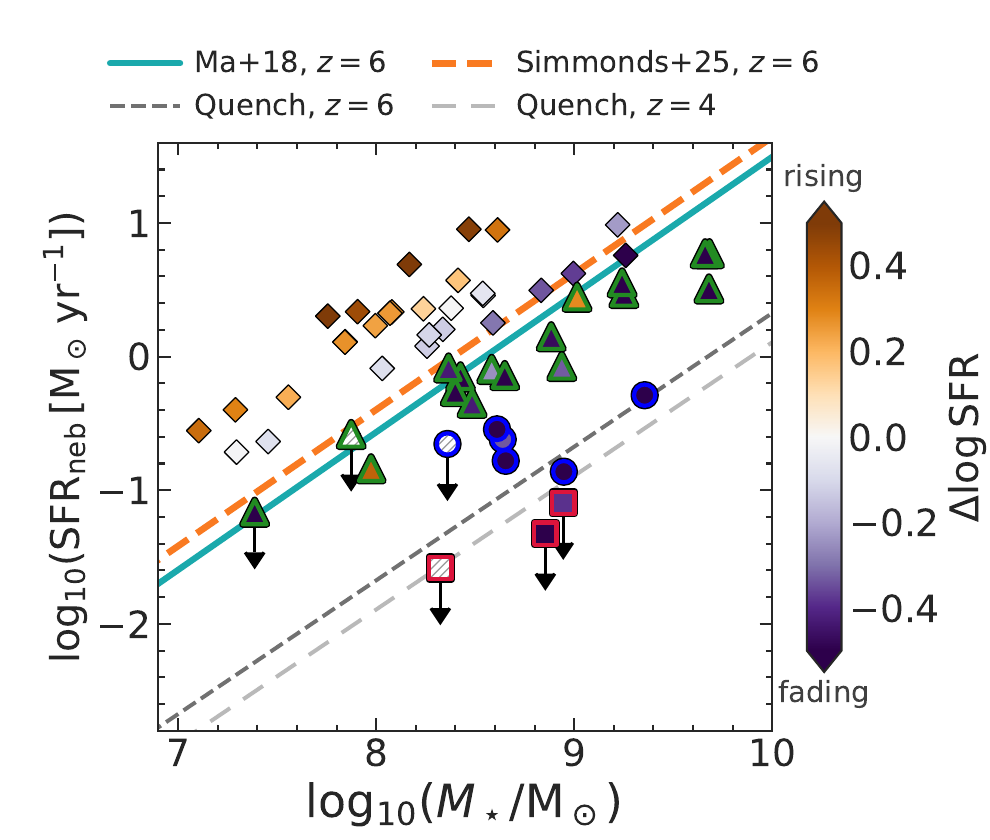}
    \caption{
\textbf{Galaxies below the SFMS have mostly faded over the last ${\sim}100$\,Myr.} The SFR--\Mstar\ plane of the $z>4$ OASIS galaxies, color-coded by the recent change in star formation.
Fill color gives $\Delta\log\mathrm{SFR}=\log\mathrm{SFR_{neb}}-\log\mathrm{SFR_{ppxf}}(100\,\mathrm{Myr})$
(Sect.~\ref{sec:sfr}), saturated at $\pm0.5$\,dex: orange for rising and purple for fading star formation.
Marker shape and edge color denote the class (OASIS Paper~I): red squares for mini-quenched, blue circles for lulling
and green triangles for nearly-MS galaxies, and black-edged diamonds for the remainder. Downward arrows are
$3\sigma$ upper limits on the nebular SFR, for which $\Delta\log\mathrm{SFR}$ is an upper limit too. Hatched symbols mark galaxies without a determined sign of $\Delta\log\mathrm{SFR}$, whose nebular upper limit lies above $\mathrm{SFR_{ppxf}}(100\,\mathrm{Myr})$ or whose $\mathrm{SFR_{ppxf}}(100\,\mathrm{Myr})$ could not be determined (see OASIS Paper~I). The lines show the
main sequences of \citet{Ma+2018} and \citet{Simmonds+2025} and the quenching threshold, evaluated at the
redshifts indicated.
}
    \label{fig:SFR-mass}
\end{figure}

\subsection{Rising and fading galaxies, and the timing of the decline}
\label{sec:tracks}

\begin{figure}
    \centering
    \includegraphics[width=\linewidth]{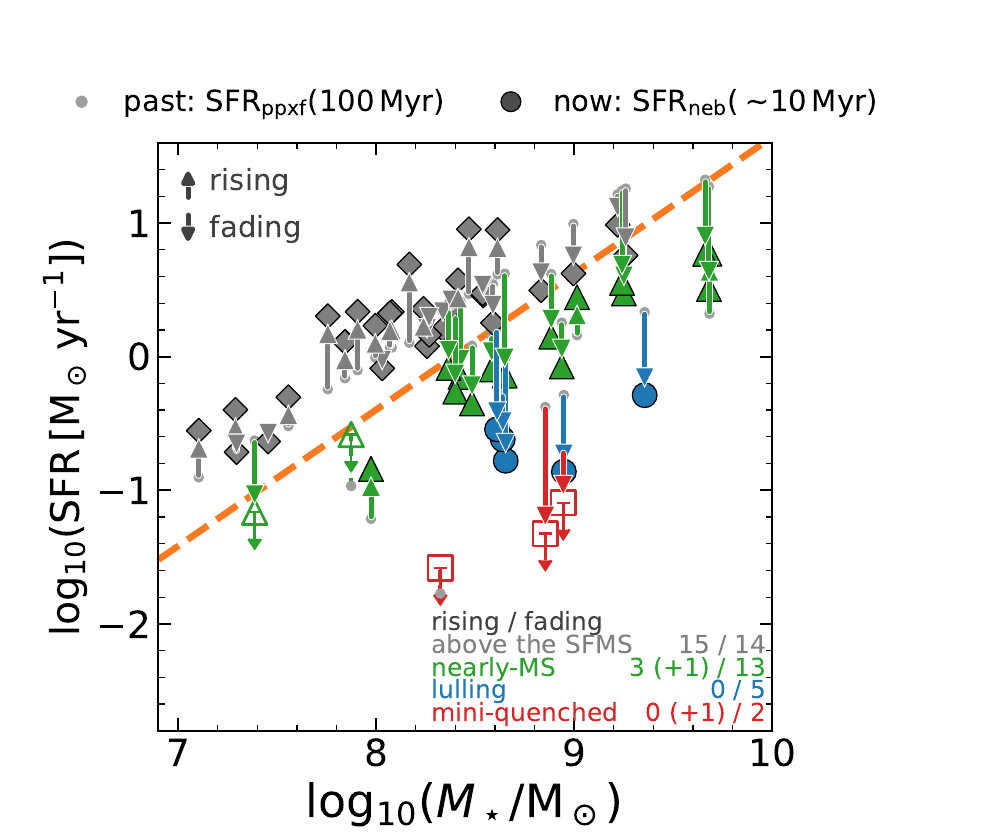}
    \caption{\textbf{Most galaxies below the SFMS have moved downward, those above it in both directions.} Where each galaxy came from and where it is ``now'', at the epoch of observation. Each arrow
    joins two SFR measurements for one galaxy: the tail, a small gray dot,
    is the $\mathrm{SFR_{ppxf}}(100\,\mathrm{Myr})$ from the \ppxf\
    star-formation history, and the head is the Balmer-line-based nebular SFR, which
    probes the last ${\sim}10$\,Myr. Symbols follow
    Fig.~\ref{fig:SFR-mass}. The tally at the lower right gives the number of rising and fading galaxies in each class; numbers in parentheses count galaxies whose upper limit on $\mathrm{SFR_{neb}}$ lies above $\mathrm{SFR_{ppxf}}(100\,\mathrm{Myr})$, so that their direction is undetermined; for these, a dotted line without an arrowhead joins the two measurements. The dashed orange line is the main sequence of
    \citet{Simmonds+2025} at $z=6$, as in Fig.~\ref{fig:SFR-mass}, shown as a fixed reference; because $\Delta_{\rm MS}$ is evaluated at each galaxy's own redshift, a galaxy can lie above this line while lying below its own main sequence.}
    \label{fig:sfr_mass_tracks}
\end{figure}

Figures~\ref{fig:SFR-mass} and \ref{fig:sfr_mass_tracks} show the 54 galaxies of the $z>4$ sample in the SFR--$M_\star$ plane, the first color-coded by $\Delta\log\mathrm{SFR}=\log\mathrm{SFR_{neb}}-\log\mathrm{SFR_{ppxf}}(100\,\mathrm{Myr})$, the second showing the same change as an arrow from $\mathrm{SFR_{ppxf}}(100\,\mathrm{Myr})$, where each galaxy sat on average over the last 100\,Myr, to $\mathrm{SFR_{neb}}$, where it sits now. Three results follow. First, the galaxies above the SFMS rise and fade equally often. Of these 29, 15 have risen ($\Delta\log\mathrm{SFR}>0$) and
14 have faded ($\Delta\log\mathrm{SFR}<0$), with a median $\Delta\log\mathrm{SFR}$ of $+0.13$\,dex; excursions are
typically small, $|\Delta\log\mathrm{SFR}|=0.25$\,dex in the median. The residual scatter (NMAD 0.32\,dex) is comparable to the combined scatter of the matched-timescale comparisons with Prospector (0.16 and 0.26\,dex, i.e., 0.30\,dex in quadrature; Sect.~\ref{sec:sfr}), so short-timescale variability within this population is not resolved. At $+0.13$\,dex, within the typical excursion of individual galaxies and measurement uncertainties, the slight positive median is consistent with a population that, on average, is neither rising nor fading; any small offset may reflect the generally rising star-formation histories expected at these redshifts and the flux-limited tiers, which favor galaxies caught while bright (OASIS Paper~I).

Second, the galaxies below the SFMS have moved predominantly downward, and
the most suppressed ones were already below it on average over the last 100\,Myr. Thirteen of seventeen
nearly-MS galaxies are fading (class median $-0.47$\,dex), and all five lulling galaxies are fading (median $-0.62$\,dex). The stellar continuum alone shows the same trend: the median ratio of the \ppxf\ SFRs over the last 10 and the preceding 90\,Myr is 0.5\,dex lower below the SFMS than above it ($p=0.02$; Sect.~\ref{sec:sfr}), mainly because of the nearly-MS galaxies. Comparing where each class is now
to where it was on average over the last 100\,Myr reveals a sequence in how long the
suppression lasted: the galaxies above the SFMS have not moved in the median
($\Delta_{\rm MS} = +0.49 \rightarrow +0.51$), the nearly-MS
galaxies were still \emph{above} the sequence and have since dropped
($+0.20 \rightarrow -0.26$), the lulling galaxies were mildly below it
($-0.35 \rightarrow -0.96$), and the mini-quenched galaxies were already
1.1\,dex below ($-1.12 \rightarrow -1.50$). The three classes below the SFMS thus form a sequence: each lay, on average over the last 100\,Myr, close to where the next class above it lies now, as expected if most of them are stages of a single decline rather than distinct populations. 

Not every galaxy needs to pass through all three stages, however: some may rise again before reaching the deepest one. Indeed, three, possibly four, of the 25 galaxies below the SFMS are rising, all nearly-MS galaxies (632746, 214959 and 116340, and possibly 214417, whose $\Delta\log\mathrm{SFR}$ is an upper limit; properties in OASIS Paper~I). Their rise is modest, only 0.3--0.45\,dex, comparable to the excursions above the SFMS, and only 632746 lay at the depth of the lulling galaxies on average over the last 100\,Myr; it is thus likely rejuvenating from a suppressed phase (spectrum in Appendix~\ref{app:rising}). Because the 100-Myr average also includes the recent rise, it may also dilute a deeper suppression, whether short or ended long enough ago for the rise to lift the average, so that the other rising galaxies could also have come up from the depth of the suppressed galaxies. Rejuvenation is, moreover, hard to identify, especially above the SFMS, where outshining by the recent burst can easily hide the traces of past suppression, such as an older stellar population and its Balmer break, so that the number of rejuvenating galaxies is a lower limit. That so few galaxies below the SFMS are rising suggests that the recovery is faster than the decline: galaxies rejuvenate too quickly to be caught often on the way up (Sect.~\ref{sec:duration}). 

Third, star formation in these galaxies does not flicker up and down on ${\sim}10$\,Myr timescales. The two measured rates can be combined into a
simple clock: if a galaxy remained on the sequence at $\mathrm{SFR_{MS}}$
until a time $t_q$ ago and has been at its present rate ($\mathrm{SFR_{neb}}$) since, the 100\,Myr
\ppxf\ average is the time-weighted mix of the two. Inverting this mix then gives
\begin{equation}
    t_q = 100\,\mathrm{Myr} \times
    \frac{1 - \mathrm{SFR_{ppxf}}(100\,\mathrm{Myr})/\mathrm{SFR_{MS}}}
         {1 - \mathrm{SFR_{neb}}/\mathrm{SFR_{MS}}}
    \label{eq:tq_main}
\end{equation}
(see Appendix~\ref{app:tq} for a derivation, discussion and visualization). For the five lulling galaxies this gives
$t_q = 18$--$81$\,Myr (median 69\,Myr). Because Eq.~\eqref{eq:tq_main}
assumes an instantaneous drop, i.e., the most abrupt decline compatible with the
two rates, these values are lower bounds on the time since the decline
began. An exponential decline, linear in $\log\mathrm{SFR}$, requires $27$--$120$\,Myr (median 112\,Myr); declines that are even more gradual are disfavored, because depletion times grow below the SFMS (Sect.~\ref{sec:mechanisms}; e.g., \citealt{Tacconi+2018}). For the mini-quenched
galaxies 168352 and 181834 the instantaneous clock gives 82 and 95\,Myr. Their nebular
rates are upper limits, but they lie so far below the SFMS that this
hardly matters: even $\mathrm{SFR_{neb}}=0$ would reduce $t_q$
only to 80 and 92\,Myr. Both values approach the 100\,Myr span of the
averaging window, and any gradual decline would place them beyond it, so
both galaxies left the sequence ${\gtrsim}100$\,Myr before observation. For
the third, 206569, the 100\,Myr average already lies below the present-day
limit, so its suppression began ${\gtrsim}95$\,Myr before observation, close to or before the start of the window.

A gradual decline lengthens the time since a galaxy left the SFMS, but the galaxy then falls below a given threshold, such as the lulling threshold, 0.6\,dex below the SFMS, or the quenching threshold, only later in the decline. For an instantaneous drop, a galaxy falls below every threshold it crosses at $t_q$, a median of ${\sim}95$\,Myr before observation for the mini-quenched galaxies, ${\sim}70$\,Myr for the lulling galaxies and ${\sim}80$\,Myr for all suppressed galaxies. In an exponential decline, $\log\mathrm{SFR}$ falls linearly from the SFMS at $t_q$ to its present value, so the galaxy crossed a threshold $\Delta_{\rm th}$ ($-0.6$ for the lulling threshold and ${\approx}-1.2$ for the quenching threshold) at a time
\begin{equation}
    t_{\rm th} = t_q\left(1-\frac{\Delta_{\rm th}}{\Delta_{\rm MS}}\right)
    \label{eq:tx}
\end{equation}
before observation, where $\Delta_{\rm MS}$ is its present offset (Appendix~\ref{app:tq}; Fig.~\ref{fig:tracks} shows one lulling and one mini-quenched galaxy). The mini-quenched galaxies then fell below the lulling threshold a median of ${\sim}95$\,Myr before observation, the lulling galaxies only ${\sim}25$\,Myr and all suppressed galaxies ${\sim}55$\,Myr before observation, and the mini-quenched galaxies fell below the quenching threshold ${\sim}40$\,Myr before observation. The speed of these declines also bears on their cause; we return to this in Sect.~\ref{sec:mechanisms}.

\begin{figure}
    \centering
    \includegraphics[width=\columnwidth]{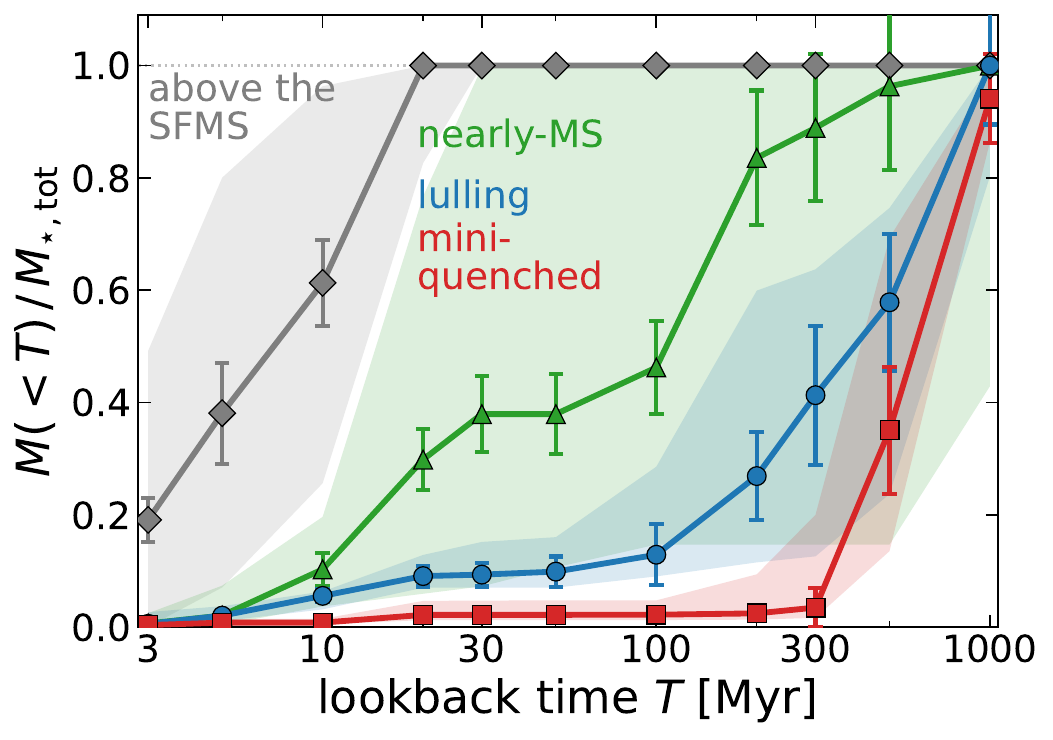}
    \caption{\textbf{The suppressed galaxies formed little of their mass in the last 100\,Myr.} The fraction of its total formed stellar mass that each galaxy
    formed within a lookback time $T$, $f(T) = M(<T)/M_{\rm tot}$, from the
    \ppxf\ star-formation histories. Curves give the median of each class: above the SFMS ($n=29$), nearly-MS ($n=17$),
    lulling ($n=5$) and mini-quenched ($n=3$). Error bars are the uncertainty on the
    median, obtained by forming the median across galaxies within each
    \ppxf\ bootstrap realization and taking the 16th--84th percentile range
    of those; the shaded bands are the 16th--84th percentile spread among
    the galaxies of each class.}
    \label{fig:mass_fraction}
\end{figure}

The cumulative mass fractions of Fig.~\ref{fig:mass_fraction} avoid the limitations of this clock. Because $f(T)$ is a ratio of two masses from
the same \ppxf\ fit rather than a rate averaged over a window, it is not diluted
as the window lengthens, and it requires neither a main-sequence relation
nor a nebular star-formation rate. According to the \ppxf\ fits, the galaxies above the SFMS have formed essentially all of their mass within the
last 20\,Myr, and the nearly-MS galaxies half of theirs within
100\,Myr. The lulling galaxies, by contrast, formed only 13 per cent of
their mass in the last 100\,Myr, and the mini-quenched galaxies just
2 per cent, a fraction that stays flat out to 300\,Myr before rising. Because the youngest populations have the lowest mass-to-light ratios, the fits are most sensitive to recent star formation, so the small mass formed within the last 100\,Myr is robust. Outshining can only hide old mass, which would lower $f(T)$ further, and the deficit does not rest on $f(T)$ alone: $\mathrm{SFR_{ppxf}}(100\,\mathrm{Myr})$ itself already lies below the SFMS (see above). For the same reason, the young ages of the galaxies above the SFMS and of the nearly-MS galaxies are lower limits (OASIS Paper~I).

Suppression is therefore sustained over several tens to ${\sim}100$\,Myr
in the lulling galaxies and over ${\gtrsim}100$\,Myr in the mini-quenched
ones, possibly up to ${\sim}300$\,Myr, the lookback time out to which their
$f(T)$ stays flat, one to two orders of magnitude longer than the nebular lines alone
can probe. The Balmer breaks corroborate this for the mini-quenched
galaxies, two of which show the strongest breaks in the entire sample
(OASIS Paper~I), which imply a continuum dominated by late-B and A stars that a galaxy
suppressed only within the last few Myr would not yet show. Their breaks are nonetheless weaker than those typical of massive quiescent galaxies at $z>4$ \citep{Baker+2025_quiescent, Weibel+2025}.

\subsection{Duration and frequency of suppressed phases: the duty cycle}\label{sec:duration}

In a steady state, the fraction of galaxies found in a given phase equals the fraction of time a galaxy spends in it. The relative numbers of galaxies above, on and below the SFMS therefore measure the duty cycle of star formation, $f_\mathrm{duty}$, which we define here as the fraction of time a galaxy spends within 0.6\,dex of the SFMS or above it. Combined with the durations of the suppressed (lulling and mini-quenched) phases, they also constrain how often galaxies are suppressed.

A formal measurement requires correcting for the OASIS selection function, which favors the suppressed population in the hand-selected tiers but disfavors it in the flux-limited ones (OASIS Paper~I), so the fractions of Sect.~\ref{sec:sample} give only a rough estimate. Taken at face value, they imply $f_\mathrm{duty}\approx0.8$ at $\log_{10}(M_\star/M_\odot)=8$--9.5, and that galaxies spend ${\sim}90$ per cent of their time above the quenching threshold.

To convert these fractions into a number of suppressed episodes, we assume that the population is in a statistical steady state and that detection does not depend on star-formation state. The suppressed fraction is then
\begin{equation}
    f_\mathrm{supp} = 1 - f_\mathrm{duty} = \frac{N\,\langle d\rangle}{T_{\rm grow}},
    \label{eq:duty}
\end{equation}
where $N$ is the number of suppressed episodes a galaxy undergoes while growing through the mass range, $\langle d\rangle$ their mean duration, and $T_{\rm grow}$ the time it spends in the range. For $T_{\rm grow}$, we integrate the growth of a galaxy along the \citet{Simmonds+2025} main sequence, $\mathrm{d}M_\star/\mathrm{d}t = (1-R)\,\mathrm{SFR_{MS}}(M_\star,z)$, with a returned mass fraction $R\approx0$--0.4. Galaxies observed at $z\approx4$--6 then take $T_{\rm grow}\approx0.5$--1.1\,Gyr to grow from $10^{8}$ to $10^{9.5}$\,M$_\odot$; these are lower limits, since time spent below the sequence slows this growth.

The mean duration $\langle d\rangle$ is not observed directly; we measure only the time elapsed within the current episode, $t_\mathrm{el}$. Because longer episodes are more likely to be caught, $\langle t_\mathrm{el}\rangle = \langle d^2\rangle/(2\langle d\rangle)$ for any distribution of durations (Appendix~\ref{app:tq}). Hence $\langle d\rangle\leq2\langle t_\mathrm{el}\rangle$ for any distribution, with equality if all episodes have the same length, since each galaxy is then caught on average halfway through its episode. Conversely, $\langle d\rangle\geq\langle t_\mathrm{el}\rangle$ if the durations are distributed no more broadly than exponentially, in which case $\langle d\rangle$ lies between $\langle t_\mathrm{el}\rangle$ and $2\langle t_\mathrm{el}\rangle$.

For an instantaneous drop, the times spent below the lulling and quenching thresholds (Sect.~\ref{sec:tracks}) give $\langle d\rangle\approx80$--160\,Myr for the suppressed and 95--190\,Myr for the mini-quenched phase. With $f_\mathrm{supp}\approx0.2$ and, for the mini-quenched phase, $f_\mathrm{mQ}\approx0.08$ (8 and 3 of the 38 galaxies at $\log_{10}(M_\star/M_\odot)=8$--9.5; Sect.~\ref{sec:sample}), a typical galaxy then undergoes $N\approx1.4$ (0.7--3) episodes of deep suppression, i.e., lulling or mini-quenched phases, while growing through this mass range, and $N\approx0.4$ (0.2--1) episodes below the quenching threshold. For an exponential decline, they give $\langle d\rangle\approx55$--110 and 40--75\,Myr, and hence $N\approx1.9$ (0.9--4) and 1.1 (0.5--2). Between the two cases, a typical galaxy undergoes one or two episodes of deep suppression while growing through this mass range, and at most about one below the quenching threshold.  

The same steady-state argument constrains how quickly galaxies recover. Only three, possibly four, of the 25 galaxies below the SFMS are rising (Sect.~\ref{sec:tracks}), and all others are declining ($\Delta\log\mathrm{SFR}\leq-0.25$).\footnote{The exception is the mini-quenched 206569, whose $\Delta\log\mathrm{SFR}$ is only an upper limit (${<}+0.19$).} Hence, galaxies below the SFMS spend of order five to seven times longer declining or suppressed than rising back towards it (21 or 22 against 3 or 4 galaxies). The comparison is cleanest within the nearly-MS range, which galaxies cross on the way down and, if they recover, on the way back up: there, 13 galaxies are fading and only 3 or 4 rising, so galaxies cross it three to four times faster upwards than downwards. Upturns in star formation are thus faster than downturns and take up only ${\sim}15$ per cent of the time galaxies spend below the SFMS.

These are order-of-magnitude estimates. The ranges reflect those of $\langle d\rangle$ and $T_{\rm grow}$, and widen further if the statistical uncertainties of the fractions are included. These estimates also rest on several assumptions: the steady state is questionable at $z>4$, where galaxies grow by large factors on ${\sim}100$\,Myr timescales; the main-sequence relation is poorly constrained at these masses; the fractions rest on three mini-quenched and five lulling galaxies subject to selection effects; and the comparison of rising and declining galaxies assumes that all suppressed galaxies eventually recover. If some never do, upturns are slower than estimated, although with expected cold-gas inflow rates per unit halo mass 50--200 times today's (Sect.~\ref{sec:mechanisms}), permanent quenching at these masses and redshifts seems unlikely. Robust constraints require larger spectroscopic samples with a well-characterized selection function.

The timescales measured for individual galaxies (Sect.~\ref{sec:tracks}) and the ages of their stellar populations (OASIS Paper~I) do not depend on the selection function and are therefore more robust than these estimates. OASIS Paper~I showed that $\Delta_\mathrm{MS}$, although set by Balmer lines that trace only ${\sim}10$\,Myr, correlates with the Balmer break and stellar ages, so it reflects changes that have persisted for ${\sim}100$\,Myr. Together, these timescales show that, in a substantial fraction of cases, galaxies do not drop from the star-forming population to below the quenching threshold within ${\sim}10$\,Myr, and some may never cross it before star formation resumes. Suppression is thus reached gradually and, once reached, can persist for ${\gtrsim}100$\,Myr. We compare these constraints with simulations in Sect.~\ref{sec:mechanisms}.

\section{Results: The ionized gas tracks the decline in star formation}\label{sec:bpt}

\begin{figure*}
    \centering
    \includegraphics[width=\textwidth]{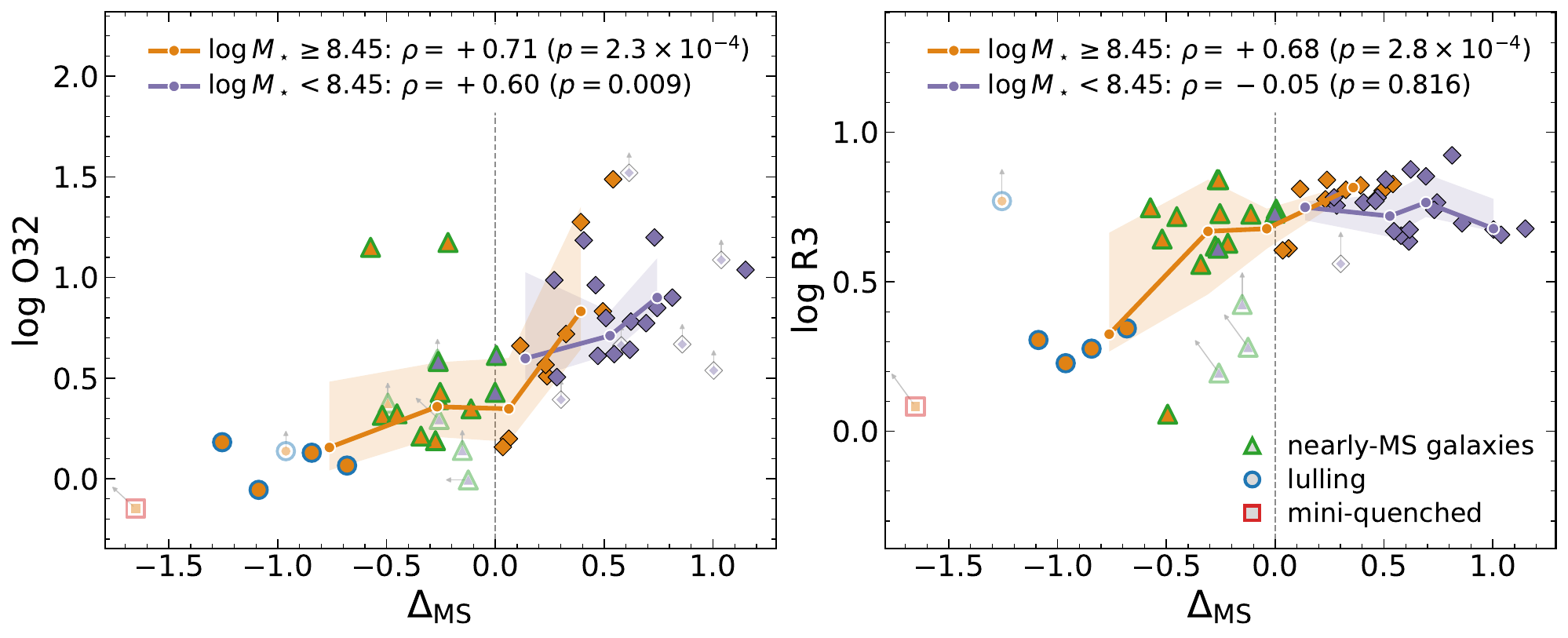}
    \caption{
    \textbf{The ionization of the gas falls with distance below the SFMS.}
    [O\,\textsc{iii}]$\lambda5007$/[O\,\textsc{ii}]$\lambda\lambda3726,3729$ (O32, left) and
    [O\,\textsc{iii}]$\lambda5007$/H$\beta$ (R3, right) against $\Delta_{\mathrm{MS}}$ for the
    $z>4$ galaxies. Fill color marks the split of the sample about the median stellar
    mass, $\log_{10}(M_\star/M_\odot)=8.45$ (orange above, purple below); marker shape and
    edge color denote the class as in Fig.~\ref{fig:SFR-mass} (legend in the right panel). Faded open symbols with arrows are limits (one line undetected). Lines and bands are the median and
    16th--84th percentile range of the detections in each mass bin, and the legend gives the
    Spearman $\rho$ with $\Delta_{\mathrm{MS}}$ in each mass bin, computed from the detections.
    }
    \label{fig:ion_dms}
\end{figure*}

\begin{table*}
\caption{Emission-line fluxes and line ratios of the suppressed galaxies and of the two rising nearly-MS galaxies of Sect.~\ref{sec:ion_dlogsfr}.}
\label{tab:lines}
\centering
\small
\setlength{\tabcolsep}{3.5pt}
\begin{tabular}{lcc ccccc ccccc}
\hline\hline
ID & $z$ & Class & [O\,\textsc{ii}] & [Ne\,\textsc{iii}] & H$\beta$ & [O\,\textsc{iii}] & H$\alpha$+[N\,\textsc{ii}] & O32 & Ne3O2 & R3 & R23 & $\Delta\log\mathrm{SFR}$ \\
 & & & 3727 & 3869 & 4861 & 5007 & & & & & & \\
\hline
168352 & 5.1 & mQ & ${<}10.3$ & ${<}9.6$ & ${<}6.1^{a}$ & $7.4\pm1.8$ & ${<}6.1$ & ${>}{-0.15}$ & -- & ${>}{0.08}$ & ${>}{0.21}$ & ${<}{-0.95}$ \\
181834 & 5.0 & mQ & ${<}14.1$ & ${<}13.4$ & ${<}9.2$ & ${<}8.2$ & ${<}9.8$ & -- & -- & -- & -- & ${<}{-0.39}$ \\
206569 & 4.0 & mQ & ${<}19.9$ & ${<}17.7$ & ${<}10.4$ & ${<}9.9$ & ${<}7.6$ & -- & -- & -- & -- & ${<}{0.19}$ \\
108487 & 4.0 & L & $43.3\pm6.3$ & ${<}18.0$ & $18.9\pm3.2$ & $38.2\pm3.2$ & $41.2\pm1.8$ & $-0.05$ & ${<}{-0.38}$ & $+0.31$ & $+0.70$ & $-0.57$ \\
114769 & 7.5 & L & ${<}7.0$ & ${<}7.0$ & $5.7\pm1.7$ & $9.7\pm2.7$ & -- & ${>}{0.14}$ & -- & $+0.23$ & ${>}{0.35}$ & $-0.74$ \\
120014 & 4.8 & L & $20.7\pm4.6$ & ${<}13.5$ & $10.9\pm2.6$ & $24.1\pm3.0$ & $35.5\pm2.2$ & $+0.07$ & ${<}{-0.18}$ & $+0.34$ & $+0.69$ & $-0.34$ \\
202438 & 5.9 & L & $22.2\pm3.1$ & ${<}8.1$ & ${<}5.7$ & $33.8\pm2.1$ & $42.0\pm2.7$ & $+0.18$ & ${<}{-0.44}$ & ${>}{0.77}$ & ${>}{1.07}$ & $-0.62$ \\
309077 & 4.7 & L & $21.9\pm4.8$ & ${<}13.1$ & $15.6\pm2.5$ & $29.5\pm2.9$ & $33.1\pm1.9$ & $+0.13$ & ${<}{-0.22}$ & $+0.28$ & $+0.59$ & $-0.77$ \\
214959 & 6.2 & nMS & $11.0\pm2.3$ & $12.5\pm2.2$ & $38.5\pm1.6$ & $164.1\pm1.7$ & $153.1\pm2.1$ & $+1.17$ & $+0.06$ & $+0.63$ & $+0.78$ & $+0.28$ \\
632746 & 6.2 & nMS & $30.2\pm2.9$ & $32.1\pm2.7$ & $76.2\pm1.9$ & $426.3\pm2.5$ & $310.3\pm2.7$ & $+1.15$ & $+0.03$ & $+0.75$ & $+0.90$ & $+0.45$ \\
\hline
\end{tabular}
\tablefoot{Observed fluxes, not corrected for dust, in units of $10^{-20}$\,erg\,s$^{-1}$\,cm$^{-2}$; non-detections are $3\sigma$ upper limits. H$\alpha$ and [N\,\textsc{ii}] are blended at the resolution of the PRISM; for 114769 at $z=7.5$, H$\alpha$ lies outside the PRISM range. Line ratios are decimal logarithms (Eqs.~\ref{eq:O32}--\ref{eq:R23}); limits follow from non-detected lines. $\Delta\log\mathrm{SFR}$ is an upper limit where $\mathrm{SFR_{neb}}$ is. Classes: mQ, mini-quenched; L, lulling; nMS, nearly-MS. $^{a}$\,A $3.7\sigma$ feature at H$\beta$ is inconsistent with Case~B given the H$\alpha$+[N\,\textsc{ii}] limit (OASIS Paper~I) and is treated as undetected; Case~B would imply H$\beta<2.1$.}
\end{table*}

We now turn to the ionized gas, using the lulling and nearly-MS galaxies to probe ISM conditions below the main sequence. All three of [O\,\textsc{ii}], H$\beta$ and [O\,\textsc{iii}] are detected in three of the five lulling galaxies (108487, 120014 and 309077); 114769 lacks [O\,\textsc{ii}] and
202438 lacks H$\beta$, so that O32 and R3 are each measured for four of them, and [Ne\,\textsc{iii}] is detected in none. Table~\ref{tab:lines} lists the fluxes and ratios of the suppressed galaxies and of the two rising nearly-MS galaxies discussed in Sect.~\ref{sec:ion_dlogsfr}.
None of the mini-quenched galaxies has both lines of any ratio detected, and only 168352 is constrained, by lower limits on O32 and R3. The weak [O\,\textsc{iii}] emission of 168352, with R3\,$>0.08$, does not constrain its ionizing source: the limit is consistent with the R3\,$\approx0.7$--0.8 of star-forming galaxies at these redshifts \citep[e.g.,][]{Cameron+2023,Sanders+2023} as well as with shocks or an AGN.

\subsection{Ionization declines below the main sequence}\label{sec:ion_sequence}

Figure~\ref{fig:ion_dms} shows that the ionization of the gas decreases steadily from the galaxies above the SFMS, through the nearly-MS galaxies, to the lulling systems. The median O32
falls from $0.78$ ($16$th--$84$th percentile $0.54$--$1.11$) for the galaxies above the SFMS to $0.39$
($0.21$--$0.66$) for the nearly-MS galaxies and $0.10$ for the lulling galaxies (range $-0.05$
to $0.18$). Each class differs significantly from the next: a Gehan test, the generalized
Wilcoxon test for data that include limits \citep{Gehan1965, Feigelson+1985}, gives
$p<0.005$, and a Mann--Whitney test on the detections alone gives consistent results
($p\leq0.006$; \citealt{MannWhitney1947}). Across the sample, O32 correlates with
$\Delta_{\mathrm{MS}}$ (Spearman $\rho=+0.69$, $p=7\times10^{-7}$; Pearson $r=+0.64$), and
does so in both mass bins ($\rho=+0.71$ above and $+0.60$ below the median mass). R3 follows the same sequence, but more
weakly ($\rho=+0.43$, $p=0.003$): the lulling galaxies have $\mathrm{R3}=0.23$--$0.34$,
against medians of $0.72$ ($0.61$--$0.75$) for the nearly-MS galaxies and $0.77$
($0.66$--$0.84$) for the galaxies above the SFMS.
Splitting in mass, we find that the R3 trend is driven by the more massive bin
($\rho=+0.68$); in the less massive bin R3 is flat with $\Delta_{\mathrm{MS}}$, but this bin
contains only six galaxies below the sequence, and only one of them (the mini-quenched 206569,
which has no measured R3) lies more than $0.5$\,dex below it.

Because all lulling galaxies lie in the high-mass bin, their low ionization could in
principle reflect the mass--metallicity relation rather than their star-formation state. This
is not the case. At fixed stellar mass, O32 still correlates with $\Delta_{\mathrm{MS}}$
(partial Spearman $\rho=+0.63$ and Pearson $r=+0.63$, $p=2\times10^{-5}$), and it does so also
at fixed mass, redshift, stellar age and stellar reddening together ($\rho=+0.62$,
$r=+0.65$), whereas at fixed $\Delta_{\mathrm{MS}}$ it shows no residual dependence on mass
($\rho=+0.19$, $r=+0.21$, $p\geq0.2$). For R3 the rank correlation at fixed mass is weaker
($\rho=+0.41$, $p=0.005$); its Pearson coefficient ($r=+0.70$) is inflated by the leverage of
the four lulling galaxies, since R3 saturates at $\approx0.8$ on and above the sequence.

Dust does not create the trend either. R3 is insensitive to dust. And although [O\,\textsc{ii}] is attenuated more than [O\,\textsc{iii}], a dust correction lowers O32 by only ${\sim}0.1$\,dex, most in the dustier galaxies above the SFMS (median nebular $E(B-V)=0.17$, against 0.06 for the lulling galaxies). It thus narrows the gap to the lulling galaxies from 0.68 to 0.66\,dex and changes the correlation of O32 with $\Delta_{\mathrm{MS}}$ by only 0.01 in $\rho$.

As noted in Sect.~\ref{sec:intro}, the ionization parameter of star-forming galaxies rises with sSFR at fixed mass out to $z\approx9$, and it also correlates with the star-formation-rate surface density \citep{reddy+2023}. In JADES at $z=1$--10, stacks of typical emission-line galaxies, which yield direct-method metallicities, show lower O32 at fixed metallicity than individual galaxies with detected auroral lines, which have higher EW(H$\beta$) and sSFR \citep{Isobe+2026}. Our results extend the link between ionization and star formation to galaxies well below the SFMS at $z>4$. They are consistent with the very low ionization of the suppressed galaxy at $z\approx8.5$ reported by \citet{Baker+2025_zap}, which rests on a tentative, $2.9\sigma$ \oii\ detection. They are also consistent with \citet{Cooper+2026}, whose low-burst-phase galaxies at
$z=3.5$--7 ($\log\mathrm{SFR_{H\alpha}/SFR_{UV}}\leq-0.3$), possibly in a lull, have lower R3 than expected from the mass--metallicity relation, which they attribute to enrichment lost in winds. Such metal loss would lower R3 below its turnover at $12+\log(\mathrm{O/H})\approx8$ \citep{Curti2020,Sanders+2024}, where galaxies of these masses and redshifts are expected to lie \citep{Nakajima+2023,Curti+2024}, but tend to raise O32, so it cannot account for the low O32 of our lulling galaxies.

At $z>10$, a low ionization state can mimic a high metallicity: the high [O\,\textsc{ii}]/H$\beta$ and low O32 from which \citet{Harikane2026} infer a near-solar metallicity for the recently quenched galaxy CEERS2-588 at $z=11$ more likely reflect, as in our lulling galaxies, a low ionization parameter. Likewise, in stacks of $z>10$ galaxies, \civ\ is detected where star formation peaked within the last 3\,Myr but not where it peaked 5--10\,Myr earlier, although C\,\textsc{iii}] persists there \citep{Roberts-Borsani2026}. Beyond $z\approx9.6$, where [O\,\textsc{iii}]\,$\lambda5007$ is redshifted out of the PRISM range, Ne3O2 can take over from O32 as a tracer of the ionization state, and hence of the recent change in star formation (Sect.~\ref{sec:ion_dlogsfr}); beyond $z\approx12.7$, where even [Ne\,\textsc{iii}] is lost, these rest-UV lines could take over in turn.

\subsection{Ionization follows the recent change in star formation}\label{sec:ion_dlogsfr}

\begin{figure}[t]
    \centering
    \includegraphics[width=\linewidth]{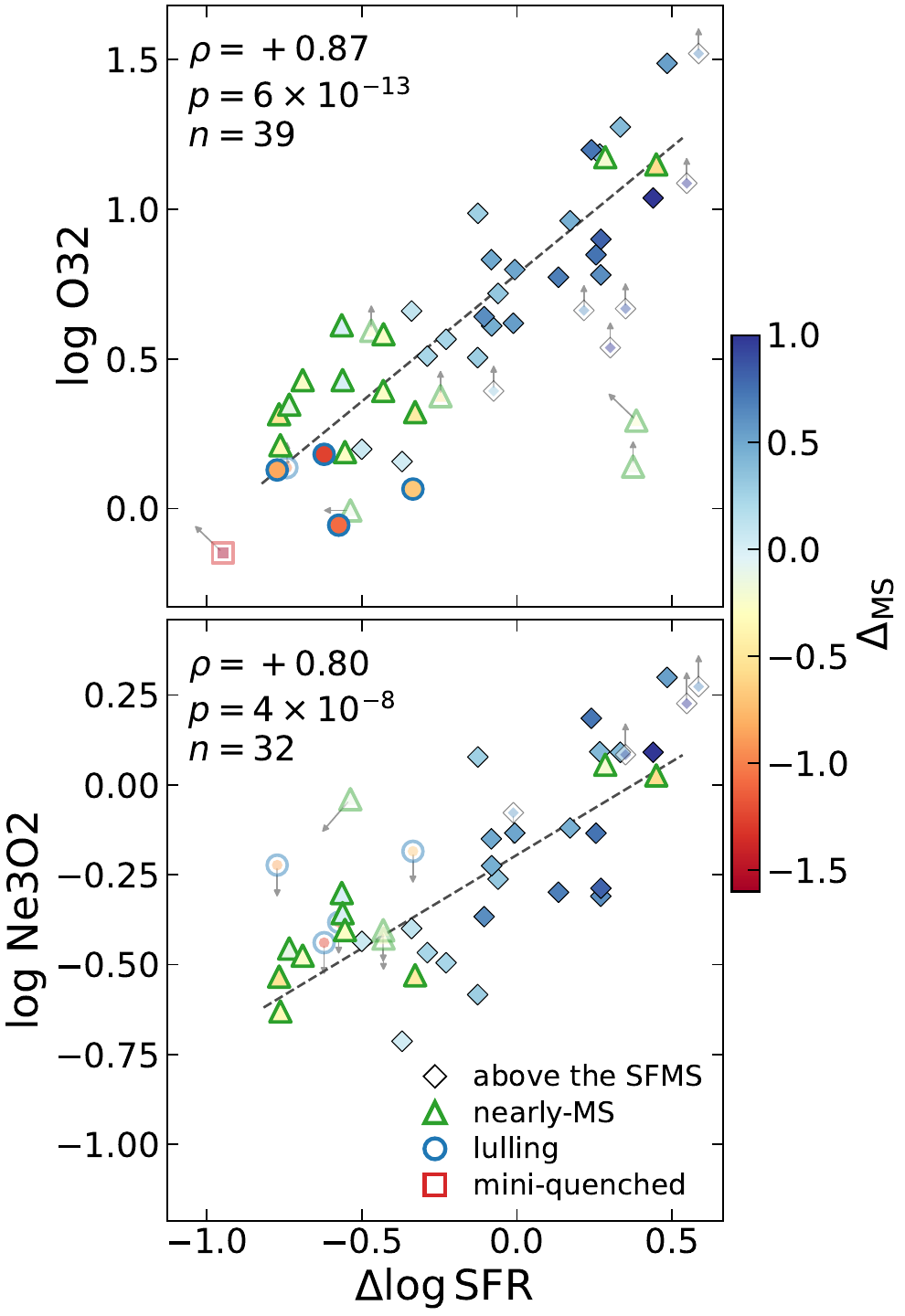}
    \caption{\textbf{The ionization of the gas tracks the recent change in star formation more closely than the distance from the SFMS.} O32 (top) and Ne3O2 (bottom) against the change of the SFR over the last ${\sim}100$\,Myr,
    $\Delta\log\mathrm{SFR}=\log(\mathrm{SFR_{neb}}/\mathrm{SFR_{ppxf}}(100\,\mathrm{Myr}))$; the
    dashed lines are least-squares fits. Marker shape and edge color give the class, as in Fig.~\ref{fig:ion_dms}, and fill color gives $\Delta_{\mathrm{MS}}$; both ratios are computed from observed fluxes.
    Faded open symbols are limits: vertical arrows mark limits on the ratios (one line
    undetected), leftward arrows upper limits on $\mathrm{SFR_{neb}}$. The correlation coefficients
    and the fits use the 39 (O32) and 32 (Ne3O2) galaxies with both lines and $\mathrm{SFR_{neb}}$ detected.}
    \label{fig:o32_dlogsfr}
\end{figure}

To identify what sets the ionization, we split the distance from the SFMS (Sect.~\ref{sec:ion_sequence}) into two terms: the change of the SFR over the last
${\sim}100$\,Myr, $\Delta\log\mathrm{SFR}$ (Sect.~\ref{sec:sfr}), and the offset of $\mathrm{SFR_{ppxf}}(100\,\mathrm{Myr})$ from the SFMS,
$\Delta_{\mathrm{MS},100}$, so that $\Delta_{\mathrm{MS}}=\Delta\log\mathrm{SFR}+\Delta_{\mathrm{MS},100}$.
We find that O32 follows the first term only. For the 39 galaxies with both O32 and $\mathrm{SFR_{neb}}$
measured, it correlates tightly with $\Delta\log\mathrm{SFR}$ (Fig.~\ref{fig:o32_dlogsfr};
$\rho=+0.87$, $p=6\times10^{-13}$; scatter $0.18$\,dex) and not with $\Delta_{\mathrm{MS},100}$ ($\rho=+0.11$, $p=0.5$). Correcting O32 for dust (Sect.~\ref{sec:ion_sequence}) leaves this correlation essentially unchanged ($\rho=+0.85$). The lulling galaxies lie at the low end of this relation, on or up to $0.43$\,dex below the fit: their ionization is as low as or lower than their decline predicts. At fixed $\Delta\log\mathrm{SFR}$, neither $\Delta_{\mathrm{MS}}$ ($\rho=+0.23$) nor stellar mass ($-0.13$) adds significant information, and sSFR at most a weak one ($+0.35$, $p=0.03$), and the
SFR itself barely predicts O32 ($\rho=+0.36$).

R3, in contrast, correlates only weakly with $\Delta\log\mathrm{SFR}$ ($\rho=+0.34$) and remains at 0.6--0.9 in nearly all galaxies on and above the sequence, whatever their recent change. At the same $\Delta\log\mathrm{SFR}$ ($-0.8$ to $-0.3$), the nearly-MS galaxies retain R3 $=0.56$--0.84, whereas the lulling galaxies have 0.23--0.34: R3 drops only far below the SFMS ($\Delta_{\mathrm{MS}}\leq-0.68$). Only the four lulling galaxies with measured R3 sample this regime, too few to tell whether R3 declines gradually or abruptly there.

Ne3O2, whose two lines lie only 140\,\AA\ apart and which is therefore nearly insensitive to dust, behaves in the same way as O32: it correlates with $\Delta\log\mathrm{SFR}$ ($\rho=+0.80$, $p=4\times10^{-8}$, for the 32 galaxies with both lines and $\mathrm{SFR_{neb}}$ detected; Fig.~\ref{fig:o32_dlogsfr}, bottom), but not with $\Delta_{\mathrm{MS}}$ at fixed $\Delta\log\mathrm{SFR}$ ($\rho=+0.06$). The correlation is strong even though [Ne\,\textsc{iii}]\,$\lambda3869$ is measured at about half the S/N of [O\,\textsc{ii}], and is affected by the stellar continuum at the Balmer break and by the unresolved He\,\textsc{i}\,$\lambda3889$ and H8 lines \citep{Tripodi+2024}. [Ne\,\textsc{iii}] is detected in none of the lulling galaxies, so this relation rests on the galaxies above and just below the SFMS.

\citet{Cooper+2026}, in contrast, find no correlation between O32 and $\log(\mathrm{SFR_{H\alpha}}/\mathrm{SFR_{UV}})$. This difference does not arise from the sample, since our correlation is unchanged without the suppressed galaxies ($\rho=+0.88$) and holds for the galaxies above the SFMS alone ($\rho=+0.82$). An H$\alpha$-to-UV ratio measured from our own spectra also correlates with O32 for these galaxies once the UV is corrected for dust: $\rho=+0.64$--0.68 with the stellar reddening of the continuum fit, and $+0.75$--0.82 with an attenuation derived from the UV slope, nearly as high as for $\Delta\log\mathrm{SFR}$; the UV slope here tracks the stellar reddening ($\rho=+0.72$) rather than the stellar age. The difference from \citet{Cooper+2026}, who measure the UV from photometry and correct it with the reddening from SED fits, may therefore lie in our measuring the UV and its attenuation directly from the spectra.

Two of the rising nearly-MS galaxies (Sect.~\ref{sec:tracks}) show most clearly that O32 follows $\Delta\log\mathrm{SFR}$ rather than $\Delta_{\mathrm{MS}}$.\footnote{The other two, 116340 and 214417, are much fainter; [O\,\textsc{ii}] is undetected in both, and their lower limits on O32 ($>0.14$ and $>0.29$) are consistent with this picture.} 632746 and 214959 (both at $z=6.2$; OASIS Paper~I and Appendix~\ref{app:rising}) lie below
the SFMS ($\Delta_{\mathrm{MS}}=-0.57$ and $-0.22$) and lay further below it on average over the last 100\,Myr
($\Delta_{\mathrm{MS},100}=-1.0$ and $-0.5$), but their star formation has since risen
($\Delta\log\mathrm{SFR}=+0.45$ and $+0.28$). The stellar continuum alone shows the same rise: their light-weighted ages of 22 and 7\,Myr indicate recent star formation, and of the 17 nearly-MS galaxies, these two have the highest ratio of the \ppxf\ SFRs over the last 10 and the preceding 90\,Myr. Their O32 ($1.15$, $1.17$) is as high as in the galaxies above the
SFMS and $0.8$\,dex above the nearly-MS median. By their
position they are nearly-MS galaxies; by the state of their gas, they are rising ones.

We interpret this through the ionization parameter $U$ and the hardness $h$ of the ionizing spectrum (the fraction of hydrogen-ionizing photons with energies above the 35.1\,eV needed to ionize O$^+$), both of which are traced by O32 \citep{Kewley+2013,Nakajima+2014}. Approximately, $\mathrm{O32}\propto h\,U$, and for an individual H\,\textsc{ii} region $U\propto(Q\,n_{\rm H}\,\epsilon^2)^{1/3}$, where $Q$ is the rate of hydrogen-ionizing photons from its stars and $n_{\rm H}$ and $\epsilon$ are the density and filling factor of the ionized gas \citep[e.g.,][]{Charlot+Longhetti2001,osterbrock+ferland2006,Gutkin+2016}. 

The ionizing photons come almost entirely from stars younger than ${\sim}10$\,Myr, and their galaxy-integrated rate is measured by $\mathrm{SFR_{neb}}$  \citep[e.g.,][]{kennicutt_star_2012}. As star formation declines, $Q$ falls and the ionizing spectrum becomes softer, because the most massive stars die first \citep[e.g.,][]{Byler2017,Eldridge2017}, so $h$ falls too. The state of the gas, in contrast, reflects the star formation of the preceding tens to ${\sim}100$\,Myr, which has consumed part of the gas, removed more through outflows \citep[e.g.,][see also Sect.~\ref{sec:mechanisms}]{Muratov+2015,Carniani+2024_outflows}, and
heated and dispersed the rest through supernovae, which begin ${\approx}3$\,Myr after stars have formed and
return ${\approx}60$ per cent of their energy and momentum by ${\approx}20$\,Myr
\citep{Faucher-Giguere2018}. All these effects grow with the star formation of the past ${\sim}100$\,Myr and lower $n_{\rm H}\epsilon^2$. 

After a decline in star formation, a weaker and softer source thus illuminates depleted,
dispersed gas, and O32 falls; after a rise from low activity, as in 632746 and 214959, the new
stars illuminate gas that little recent star formation has used up or dispersed, and O32 is high.
$\Delta\log\mathrm{SFR}$ then measures the balance between the current photon supply and the
earlier star formation that depleted and dispersed the gas.

These results agree with simulations in
which the gas follows the burst cycle: in SPHINX20, galaxies with
$\mathrm{SFR_{10}}<\mathrm{SFR_{100}}$ have much lower H\,\textsc{ii}-region densities, and their ISM
is disrupted by the supernovae of the previous burst \citep{Katz+2023}.  In the SPHINX20
comparison of \citet{Witten+2025a}, low O32 likewise coincides with low
$\mathrm{SFR_{10}}/\mathrm{SFR_{100}}$, although their rejuvenating galaxy A2744-YD4 has low O32
(O32 $=0.12$) despite its recent burst, which they attribute to an ISM enriched and disrupted by a preceding, longer burst. In SPHINX20, O32 also correlates with the sSFR and peaks while the ionizing population is younger than ${\sim}3.5$\,Myr \citep{Choustikov+2023}, consistent with O32 following the current ionizing photon supply and its hardness. In THESAN-ZOOM, galaxy-scale outflows peak a median ${\sim}21$\,Myr after the star formation \citep{McClymont+2025}, and after a burst feedback clears the dense gas around the young stars and opens low-density channels in the ISM \citep{Summerfield+2026}.

\subsection{Lulling galaxies in the R23--O32 diagram}\label{sec:r23o32}

\begin{figure}
    \centering
    \includegraphics[width=\linewidth]{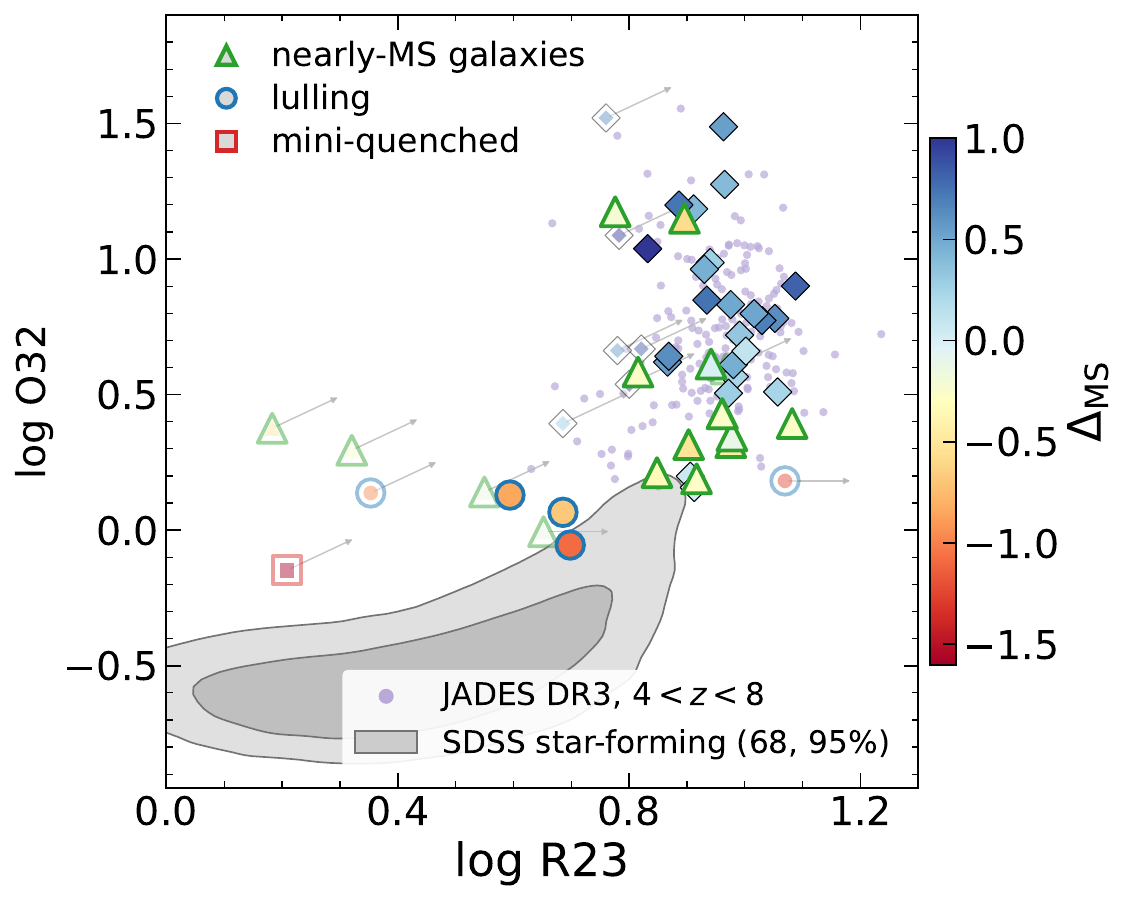}
    \caption{
    \textbf{The lulling galaxies lie between the high-redshift population and local
    star-forming galaxies in O32--R23.}
    O32 against R23 (Eqs.~\ref{eq:O32} and \ref{eq:R23}) for the $z>4$ galaxies. Fill color gives $\Delta_{\mathrm{MS}}$; marker shape and edge color
    denote the class as in Fig.~\ref{fig:SFR-mass}. Faded open symbols with arrows are limits (one line undetected). Small purple points are the 182 galaxies at $4<z<8$ from the JADES DR3 prism
    line-flux catalog \citep{DEugenio+2025_JADESDR3} with [O\,\textsc{ii}], H$\beta$ and [O\,\textsc{iii}] measured. Gray contours enclose 68 and 95 per cent of the SDSS DR7 star-forming galaxies
    \citep[MPA-JHU catalog;][]{abazajian+2009, brinchmann+2004}, selected with the
    \citet{Kauffmann2003} line and with signal-to-noise ratio $>5$ in all lines. Ratios are of observed fluxes for all three samples.
    }
    \label{fig:r23o32}
\end{figure}

Figure~\ref{fig:r23o32} places our OASIS galaxies on the R23--O32 diagram. As found for other JWST samples \citep[e.g.,][]{Cameron+2023, Mascia+2023, Tang+2023}, the galaxies above the SFMS are offset from the local SDSS locus to higher O32 at fixed R23. The three lulling galaxies with both ratios measured instead occupy the region between this
high-redshift population and local star-forming galaxies, with the lowest O32 of all such galaxies and lower R23 (median $0.69$, against $0.93$ and $0.97$ for the nearly-MS galaxies and those above the SFMS). Such line ratios are essentially absent from spectroscopic samples at these redshifts that were not selected for suppressed galaxies: of 182 galaxies at $4<z<8$ in the JADES
prism catalog \citep{DEugenio+2025_JADESDR3}, which occupy the same locus as our galaxies above the SFMS (median O32 $=0.72$ and R23 $=0.97$, against $0.78$ and $0.97$ for ours), only two have O32 $<0.2$, and none
combines O32 $<0.2$ with R23 $<0.75$. That catalog includes only lines detected at high signal-to-noise ratio and so favors strong emitters, but this is precisely the point:
galaxies in a lull are faint in their lines, and are missed by line-selected samples.

\subsection{Ionizing sources and the need for grating spectroscopy}\label{sec:ion_source}

The line ratios used in this section trace the ionization state of the gas
but do not identify its ionizing source. At the high ionization parameters and low metallicities common at
$z\gtrsim4$, optical strong-line ratios are often set more by $\log U$ and the gas-phase metallicity
than by the shape of the ionizing spectrum, so that stellar and black-hole-accretion-powered
ionization become degenerate \citep[e.g.,][]{Cleri+2025, Maiolino2024, Mazzolari+2024}. Local demarcations between star formation and
AGN therefore do not transfer directly to $z\gtrsim4$, where purely star-forming galaxies can
occupy the local AGN locus
\citep[e.g.,][]{Nakajima+2023, Sanders+2023, Cameron+2023, Trump+2023} and AGN can populate the
star-forming region \citep{Kocevski+2023, Uebler+2023, Juodzbalis+2026}. With the PRISM data, a contribution from Type~2 AGN or shocks can be neither confirmed nor ruled out for individual galaxies (Appendix~\ref{app:bpt}), and at the resolution of the PRISM a broad H$\alpha$ component from a Type~1 AGN cannot be separated from blended [N\,\textsc{ii}] emission (Sect.~\ref{sec:mechanisms}).

A more detailed characterization of the suppressed galaxies requires deep medium-resolution grating spectroscopy, to measure low-equivalent-width lines, deblend closely spaced transitions such as H$\alpha$ and \nii, search for multiple kinematic components, and measure electron densities from the resolved [S\,\textsc{ii}] and, where possible, [O\,\textsc{ii}] doublets, which would show whether a lower gas density contributes to the low ionization. Such observations can help identify the dominant ionizing sources and the mechanisms that suppress star formation in these galaxies. This is the primary motivation for our JWST Cycle~5 program LULLs (GO~10149; PI: T.~J.~Looser, co-PI: F.~D'Eugenio), which will obtain NIRSpec G235M + G395M grating spectroscopy (18 + 60 hours) of such galaxies.

\section{Discussion: Evidence for feedback-driven suppression}\label{sec:mechanisms}

OASIS Paper~I and the preceding sections establish four properties of the suppressed galaxies. First, they show little stellar reddening, with median $E(B-V)_\star\approx0.1$ against 0.20 for the rest of the sample (OASIS Paper~I). Second, they are not caught in a brief dip: they formed little of their mass in the last 100\,Myr (Fig.~\ref{fig:mass_fraction}), and their star formation declined tens to ${\gtrsim}100$\,Myr before observation (Sect.~\ref{sec:tracks}). Third, such phases are common: a typical galaxy undergoes one or two such phases while growing through $10^{8}$--$10^{9.5}$\,M$_\odot$ (Sect.~\ref{sec:duration}). Fourth, their gas is in a low ionization state that tracks the decline of their star formation (Sect.~\ref{sec:bpt}). Here we use these results to ask what suppresses their star formation: whether the galaxies still contain gas, whether their declines are too fast for the gas to have been consumed, how they compare with simulations, what could sustain the suppression, and whether an AGN contributes.

\paragraph{The galaxies may have lost their gas.}
Given the low stellar reddening of these galaxies, and assuming this reflects a low dust mass, typical dust-to-gas ratios would imply a correspondingly low molecular
gas mass. This inference comes with three caveats. First, the dust-to-gas ratio drops steeply and with large scatter
below $\sim0.1\,Z_{\odot}$ \citep[e.g.,][]{Remy-Ruyer+2014, Li+2019}, which is
close to the metallicities expected for low-mass galaxies at these redshifts
\citep[e.g.,][]{Nakajima+2023, Curti+2024,
Sanders+2024, Chemerynska+2024}. Second, dust may
be preferentially removed by radiation pressure without a comparable
removal of gas \citep[e.g.,][]{Ferrara+2024}. Third, supernova shocks can destroy dust without removing the gas \citep{Hirashita+2002,Slavin+2015}. If, however, these systems follow
the dust-to-gas scaling of local galaxies, their low gas content would favor a
scenario in which star formation is suppressed because the galaxy has been depleted of its fuel, either through gas consumption or through ejective feedback, rather than one in which the gas is
merely heated or redistributed by turbulence while remaining in the system.
Metal-poor galaxies with rising star formation below the SFMS at $\log_{10}(M_\star/M_\odot)\leq8$ point the same way: \citet{Isobe+2026} interpret them as emerging from a gas-poor, mini-quenched phase.

A direct test is the molecular gas itself, which can be constrained through [C\,\textsc{ii}]\,$158\,\mu$m, the only practical tracer at these metallicities and redshifts \citep[e.g.,][]{Zanella+2018, DEugenioC+2023, Valentino+2026}. In dormant galaxies at $\log_{10}(M_\star/M_\odot)>9.5$ and $z>3$, archival [C\,\textsc{ii}] and CO limits already place the molecular gas fractions 0.3--2\,dex below main-sequence values \citep{Xiao+2026}. At our masses the measurement remains demanding, however: constraining the gas-to-stellar mass ratio to ${\lesssim}0.5$--1 for an individual unlensed object of $M_{\star}\sim5\times10^{8}\,M_{\odot}$ at $z\approx4$--8 would require ${\sim}10$--100\,h with ALMA (for a $3\sigma$ limit, an unresolved line of 100\,km\,s$^{-1}$ FWHM and $\alpha_{\rm [CII]}\approx30\,M_\odot/L_\odot$; \citealt{Zanella+2018}), and lower gas fractions become accessible only by stacking ${\sim}10$ systems or through strong gravitational lensing ($\mu \gtrsim 5$).

\paragraph{The declines are too fast for gas consumption alone.}
If the suppressed galaxies are indeed poor in gas, the missing gas has either been consumed by star formation or expelled by feedback; if they are not, feedback may instead keep the gas from forming stars. The speed of the decline in star formation distinguishes consumption from both feedback channels: starvation, the passive exhaustion of the remaining gas without replenishment, lowers the SFR only on gas depletion timescales, whereas feedback that removes the gas or keeps it from forming stars can suppress star formation much faster. We therefore test the null hypothesis that consumption alone accounts for the measured declines (Sect.~\ref{sec:tracks}) by comparing two e-folding times: the longest compatible with the observed declines in star formation, and the shortest that passive exhaustion can produce. If even the fastest passive decline is slower than the slowest observed one, consumption alone cannot account for the declines. At each step below, we make the choice that favors consumption.

To compare the two, we describe the gas reservoir with a star-formation law $\mathrm{SFR}=A\,M_{\rm gas}^{k}$ and assume that no gas is accreted during the decline, since any inflow would slow it. The gas is then consumed by star formation, with a returned mass fraction $R$, and, in the feedback alternative, by outflows of mass-loading factor $\eta$,
\begin{equation}
  \frac{\mathrm{d}M_{\rm gas}}{\mathrm{d}t} = -(1-R+\eta)\,\mathrm{SFR},
  \label{eq:gas_balance}
\end{equation}
so that the SFR declines as
\begin{equation}
  \frac{\mathrm{d}\,\mathrm{SFR}}{\mathrm{d}t} = k A\,M_{\rm gas}^{k-1}\,\frac{\mathrm{d}M_{\rm gas}}{\mathrm{d}t}
  = -\frac{\mathrm{SFR}}{\tau},
  \qquad
  \tau = \frac{t_{\rm depl}}{k\,(1-R+\eta)},
  \label{eq:tau_eta}
\end{equation}
where $t_{\rm depl}=M_{\rm gas}/\mathrm{SFR}=M_{\rm gas}^{1-k}/A$ is the depletion time. Depletion times grow below the main sequence \citep[e.g.,][]{Genzel+2015, Saintonge+2017, Tacconi+2018, Tacconi2020}, which implies $k>1$: $t_{\rm depl}$ then grows as the gas is consumed, so the decline slows with time and is fastest at its onset.\footnote{A sublinear law ($k<1$), whose decline steepens towards exhaustion, would reproduce the observed SFR ratios (Eq.~\eqref{eq:tau_decline}) only several hundred Myr after the decline began, a large fraction of the age of the Universe at $z=4$--7.5.}

The nebular and \ppxf\ rates measure the mean SFR over the last ${\sim}10$ and 100\,Myr (Sect.~\ref{sec:sfr}), so their ratio traces how fast star formation has declined. We describe the observed declines as exponential, the decline of Eq.~\eqref{eq:tau_eta} for a constant depletion time ($k=1$), now fitted to the ratio of the two rates alone, without assuming that the decline began on the SFMS (cf. Sect.~\ref{sec:tracks}). This is the conservative choice: it is the slowest decline that reproduces the observed ratios, since any other shape must fall faster than $e^{-t/\tau}$ somewhere within the window, so the e-folding times derived from it are upper limits on the e-folding time where the actual decline is steepest. With $\mathrm{SFR}\propto e^{t_{\rm lb}/\tau}$ in lookback time $t_{\rm lb}$, the mean SFR over the last $T$ is $\langle\mathrm{SFR}\rangle_T\propto\tau\,(e^{T/\tau}-1)/T$, and the ratio fixes $\tau$ through
\begin{equation}
  \frac{\mathrm{SFR_{neb}}}{\mathrm{SFR_{ppxf}}(100\,\mathrm{Myr})} =
  \frac{\langle\mathrm{SFR}\rangle_{10\,\mathrm{Myr}}}{\langle\mathrm{SFR}\rangle_{100\,\mathrm{Myr}}}
  = 10\,\frac{e^{10\,\mathrm{Myr}/\tau}-1}{e^{100\,\mathrm{Myr}/\tau}-1}.
  \label{eq:tau_decline}
\end{equation}
This gives $\tau=32$--66\,Myr (median 39\,Myr) for the lulling galaxies, and ${<}27$ and ${<}58$\,Myr for the mini-quenched galaxies 168352 and 181834. For the third, 206569, the 100-Myr average already lies below the nebular limit, so its decline began close to or before the start of the window and $\tau$ is unconstrained (Sect.~\ref{sec:tracks}).

For the fastest decline that passive exhaustion can produce, we take the depletion time of the molecular gas from the empirical relation of \citet{Tacconi+2018},
$t_{\rm depl}=1.23\,\mathrm{Gyr}\,(1+z)^{-0.62}\,\delta_{\rm MS}^{-0.44}\,(M_\star/10^{10.7}M_\odot)^{0.09}$,
where $\delta_{\rm MS}$ is the offset from the main sequence of \citet{speagle+2014}. It is calibrated only for $\log_{10}(M_\star/M_\odot)\geq9$, $z\leq4$ and $\delta_{\rm MS}\geq10^{-1.3}$, so at our masses and redshifts it is an extrapolation in all three quantities, and it describes the population rather than the track of an individual galaxy. At fixed mass and redshift $\delta_{\rm MS}\propto\mathrm{SFR}$, so the relation implies $M_{\rm gas}\propto\mathrm{SFR}^{0.56}$, i.e. $k=1/0.56\approx1.8$. Since $k>1$, we conservatively evaluate the decline at its onset, when the galaxies were at most on the main sequence (their 100-Myr average already lies below it; Sect.~\ref{sec:tracks}) and the depletion time was shortest. With the main-sequence depletion time ($\delta_{\rm MS}=1$), $t_{\rm depl}\approx0.2$--0.3\,Gyr, and neglecting the gas returned by stellar mass loss ($R=0$), which would slow the decline further, passive exhaustion ($\eta=0$) gives $\tau\approx120$--180\,Myr; the relations of \citet{Genzel+2015} and \citet{Scoville+2017} give similar or longer values.

Despite our assumptions favoring the null hypothesis of gas consumption, these passive e-folding times are two to five times longer than the upper limits measured for the lulling galaxies, and ${\gtrsim}3$ and ${\gtrsim}6$ times longer than those of 181834 and 168352. The lulling galaxies instead require outflows with $\eta\approx1$--4 (181834 and 168352: ${\gtrsim}2$ and ${\gtrsim}5$), or $\eta\approx7$--23 with the depletion times at the observed offsets below the main sequence ($\delta_{\rm MS}\approx10^{-1.2}$--$10^{-1.7}$, where the relation gives $t_{\rm depl}\approx0.7$--1.7\,Gyr). Because the measured e-folding times are upper limits, these values of $\eta$ are lower limits. Alternatively, the gas may be kept from forming stars.

The declines are therefore too rapid for the galaxies to be simply running out of fuel, and the null hypothesis is disfavored. The assumption of no inflow is, moreover, generous, because starvation is difficult to achieve at these redshifts in the first place. Halos of the masses expected for our galaxies lie far below the mass above which accreted gas is shock heated, so that they are fed by cold, filamentary streams that reach the galaxy directly \citep{Keres+2005, DekelBirnboim2006}, and the inflow rate per unit halo mass scales as $(1+z)^{5/2}$ \citep{Dekel+2013}, 50--200 times its present-day value at $z=4$--7.5. Stopping this supply itself requires feedback, which can heat or push back the inflowing gas \citep{Faucher-Giguere+2011}.

Both feedback channels, removing the gas and keeping it from forming stars, have
observational support at higher masses. Feedback-driven outflows are seen directly in massive quiescent galaxies (Sect.~\ref{sec:intro}). The ejective picture is,
however, contested: a massive quiescent galaxy at a comparable redshift retains a molecular gas fraction of ${\approx}20\%$, with further [C\,\textsc{ii}] emission extending to circumgalactic scales, which in that case favors
preventative quenching: the gas is retained but kept from forming stars
\citep{Valentino+2026}. Whether either feedback mode
extends to the low-mass regime probed here remains an open question.

\paragraph{Comparison with simulations.}
The (conservative) mass-loading factors we require, $\eta\gtrsim1$--5, or up to ${\approx}23$ for the depletion times at the observed SFMS offsets, are comparable to or up to a few times above those of simulations with explicit stellar feedback, which at $M_\star\approx10^8$--$10^{9.5}$\,M$_\odot$ and $z\lesssim4$ reach $\eta\approx5$--20 in \textsc{Fire} and ${\approx}1$--5 in \textsc{Fire}-2, rising to ${\sim}100$ only at $M_\star\lesssim10^{6}$\,M$_\odot$ \citep{Muratov+2015, Pandya+2021}.
Stellar feedback can therefore plausibly produce declines as rapid as we observe, and it drives most episodes of temporary quiescence in simulated low-mass galaxies \citep{Gelli+2025} through cycles of ejection and re-accretion of the ISM \citep[e.g.,][]{McClymont+2025}.
Interrupted gas accretion and galaxy interactions contribute as well in simulations \citep{Dome2023}, but are unlikely to dominate in our galaxies: their star formation falls faster than an interrupted supply alone would allow, and they are isolated (OASIS Paper~I).

How long suppressed phases last is a key test of models of bursty star formation, and ours last as long as or longer than most simulations predict.
In the zoom-in simulations analyzed by \citet{Dome2023}, at $z=4$--8, galaxies spend a median of 22 and 42\,Myr at least 0.6\,dex below the main sequence, our lulling threshold, in \textsc{Vela} and \textsc{FirstLight}, respectively, with rare excursions to ${\sim}100$--220\,Myr.
Because longer episodes are more likely to be caught (Sect.~\ref{sec:duration}), these distributions predict that galaxies observed in such a phase crossed this threshold ${\sim}20$--50\,Myr earlier. Our suppressed galaxies crossed it a median of ${\sim}55$--80\,Myr ago, for declines between exponential and instantaneous (Sect.~\ref{sec:tracks}), comparable to or up to a few times longer than predicted.
The three mini-quenched galaxies have spent a median of ${\sim}95$\,Myr more than 0.6\,dex below the main sequence and ${\sim}40$--95\,Myr below the quenching threshold, for exponential to instantaneous declines (Sect.~\ref{sec:tracks}), whereas in \textsc{Sphinx}$^{20}$, mini-quenched galaxies remain so for less than 20\,Myr on average \citep{Faisst+2024}.

Suppressed phases as long as those in OASIS also appear in other simulations, but are defined less strictly, at lower masses or only in very few systems: in \textsc{Thesan-Zoom}, episodes below the main sequence, at any depth, last a median of ${\sim}80$\,Myr at $M_\star>10^8$\,M$_\odot$ \citep{McClymont+2025}; in \textsc{Megatron} at $z\geq8.5$, star formation in two mini-quenched systems dropped sharply 70--80\,Myr before observation \citep{Katz+2025}, and in the runs with the strongest feedback, star formation in quenched dwarfs has ceased for a median of ${\sim}135$\,Myr, but only at $M_\star\lesssim10^{8}$\,M$_\odot$ \citep{Attar+d2026}.
In \textsc{Serra}, quiescent galaxies dominate below $M_\star\approx10^{8.3}$\,M$_\odot$, and the fraction of time with ongoing star formation rises from ${\approx}0.6$ at $10^{7.5}$\,M$_\odot$ to ${\approx}0.99$ at $M_\star\geq10^{9}$\,M$_\odot$ \citep{Gelli+2023}, bracketing our rough estimate of ${\sim}0.9$ above the quenching threshold at $10^{8}$--$10^{9.5}$\,M$_\odot$ (Sect.~\ref{sec:duration}).

\paragraph{Sustaining the suppression is harder than starting it.}
The difficulty lies not in starting the suppression but in sustaining it. Supernovae begin to explode within ${\approx}3$\,Myr \citep{Faucher-Giguere2018}, but their galaxy-scale effect on star formation
builds up over ${\gtrsim}30$\,Myr \citep{Gelli+2023}, comparable to the e-folding times we
infer. Once the gas has been expelled, however, it returns quickly in the simulations, whose
suppressed phases are at most as long as ours and occur mainly at lower masses (see above). Either supernova feedback keeps the ejected gas from returning for longer than the
simulations predict, or one or more additional mechanisms are at work, as has also been suggested from the
abundance of temporarily quiescent galaxies \citep{Gelli+2023, Gelli+2025}.

One candidate for an additional mechanism is radiation pressure from young massive stars. Galaxies with specific SFRs above ${\approx}25\,\mathrm{Gyr^{-1}}$ become
super-Eddington, and radiation pressure on dust drives outflows that clear the dust from the
star-forming regions and displace part of the gas to larger radii \citep{Ferrara+2024}. \citet{Gelli+2023,Gelli+2023b} also invoked such winds, because supernova feedback alone cannot
truncate star formation within a few Myr of a burst, as required for the mini-quenched galaxy
of \citet{Looser+2024Nat}. Four of our rising galaxies above the SFMS, at $\log_{10}(M_\star/M_\odot)=7.8$--8.5, currently exceed this threshold, but whether they will enter a suppressed phase shortly after observation is unknown.
If the bursts that preceded the suppression of our mini-quenched and lulling galaxies reached such specific SFRs, this would link the rapid declines to their low dust content, but our SFHs do not resolve these bursts finely enough to test this.

Another candidate is feedback from
accreting massive black holes. In the \textsc{Flares} simulations, AGN feedback is the main driver of passivity at high redshift in galaxies of $M_\star\gtrsim5\times10^{9}$\,M$_\odot$, acting through reduced gas fractions \citep{Lovell+2023}. Zoom-in simulations of dwarf galaxies show that it can operate at
low masses too, in two different ways: efficient AGN can drive the gas out and then keep the
galaxy quiescent by heating its circumgalactic medium \citep{koudmani+2022}, or they can heat
the central gas and push it outwards, suppressing star formation while the total gas mass is preserved
\citep{Arjona-Galvez+2024}.

\paragraph{An AGN contribution cannot be excluded.}
Our spectra leave this possibility open. In all four lulling galaxies with H$\alpha$ coverage (309077, 120014,
202438 and 108487), the H$\alpha$ profile is not reproduced by a single Gaussian
(OASIS Paper~I), as expected for either a broad component or strong
[N\,\textsc{ii}] emission. One of them, 202438, is also unusual: H$\alpha$ is detected but
H$\beta$ is not, and the implied Balmer decrement, $E(B-V)_{\rm neb}\gtrsim0.9$, is far above its
stellar reddening of 0.14. Dust around the ionized gas could explain this, but so could
[N\,\textsc{ii}] emission blended with H$\alpha$ at PRISM resolution, or a broad component
whose H$\beta$ counterpart falls below our detection limit. The narrow-line ratios cannot
settle the question. The low ionization
state of the lulling galaxies (Sect.~\ref{sec:bpt}) argues against a luminous AGN dominating
the ionization of their gas, but a weaker or obscured one cannot be excluded, because at $z\gtrsim4$ AGN and star-forming galaxies overlap in the line-ratio diagrams (Sect.~\ref{sec:ion_source} and Appendix~\ref{app:bpt}). The mini-quenched galaxies tell a different
story: without any Balmer emission they cannot host a luminous, currently accreting, unobscured AGN
(OASIS Paper~I), so that if an AGN suppressed their star formation, it has since
faded. A lull in star formation might then coincide with a still active black hole, and a
quenched phase with one that has already switched off. Medium-resolution spectra of the
H$\alpha$ profiles will show whether the lulling galaxies host such active black holes.

\section{Conclusions}\label{sec:Summary}

JWST has shown that star formation in low-mass galaxies at $z>4$ is bursty, and OASIS Paper~I found that the galaxies caught in the quiet phases of this cycle are common and old. A single snapshot, however, cannot tell how they got there: how fast their star formation declined, what drove the decline, and how their gas and ISM properties changed along the way. These are the most direct tests of feedback models. 

We have investigated these questions with deep NIRSpec/PRISM spectra of 54 galaxies at $z>4$ from the JWST program OASIS (GO~5997). Of these, 29 lie above the star-forming main sequence (SFMS) and 17 are nearly-MS galaxies, less than 0.6\,dex below it; five are lulling, at least 0.6\,dex below it but above the quenching threshold, a specific SFR of $0.2/t_{\rm obs}$; and three are mini-quenched, below this threshold. We refer to the lulling and mini-quenched galaxies together as suppressed. 
Comparing the Balmer-line SFRs with those averaged over the last 100\,Myr from full spectral fitting, through their logarithmic ratio $\Delta\log\mathrm{SFR}$, places each galaxy on its recent track through the SFR--$M_\star$ plane. Both rates agree with independent photometric (Prospector) estimates on matched timescales, so that most of the scatter in $\Delta\log\mathrm{SFR}$ is physical (Sect.~\ref{sec:sfr}). The emission lines then trace how the state of the gas responds to this change in star formation. Our main conclusions are as follows:

\begin{itemize}

\item \emph{Galaxies below the SFMS are fading, not flickering.} Galaxies above the SFMS rise and fade equally often: 15 of 29 have risen ($\Delta\log\mathrm{SFR}>0$) and 14 have faded ($\Delta\log\mathrm{SFR}<0$), but have moved very little, with a median $|\Delta\log\mathrm{SFR}|$ of only 0.25\,dex. Below the SFMS, in contrast, star formation has declined in 13 of 17 nearly-MS galaxies and in all five lulling galaxies, with class medians of $\Delta\log\mathrm{SFR}=-0.47$ and $-0.62$\,dex, and the most suppressed, the lulling and mini-quenched galaxies, were already below the SFMS on average over the last 100\,Myr, by a median of 0.35 and 1.1\,dex. The three classes below the SFMS thus form a sequence, as expected if most of them are stages of a single decline rather than distinct populations (Sect.~\ref{sec:tracks}).

\item \emph{The suppression is not a brief dip.} The lulling galaxies formed only 13 per cent of their mass within the last 100\,Myr, and their Balmer-line and 100-Myr SFRs imply that their decline began 18--81\,Myr (median 69\,Myr) before observation for an instantaneous drop from the SFMS, or ${\sim}110$\,Myr for an exponential decline; declines that are more gradual are disfavored, because depletion times grow below the SFMS (Sect.~\ref{sec:mechanisms}). The mini-quenched galaxies left the SFMS ${\gtrsim}100$\,Myr ago and formed only 2 per cent of their mass within the last 100--300\,Myr (Sect.~\ref{sec:tracks}).

\item \emph{Suppressed phases are long and common.} In an order-of-magnitude estimate, galaxies at $\log_{10}(M_\star/M_\odot)=8$--9.5 spend ${\sim}20$ per cent of their time in a lulling or mini-quenched phase, and these episodes last ${\sim}55$--160\,Myr, as inferred from the elapsed times. A typical galaxy therefore undergoes one or two such episodes while growing from $10^{8}$ to $10^{9.5}$\,M$_\odot$ (Sect.~\ref{sec:duration}). The time elapsed since our suppressed galaxies fell 0.6\,dex below the SFMS is comparable to or up to a few times longer than in the \textsc{Vela} and \textsc{FirstLight} zoom-in simulations, and the mini-quenched galaxies have spent ${\sim}40$--95\,Myr below the quenching threshold, against less than 20\,Myr on average in \textsc{Sphinx}$^{20}$ (Sect.~\ref{sec:mechanisms}). If suppressed galaxies eventually recover, as expected given their high gas inflow rates, those below the SFMS spend five to seven times longer declining or suppressed than rising back towards it, and cross the nearly-MS range three to four times faster upwards than downwards (Sect.~\ref{sec:duration}).

\item \emph{The ionized gas tracks the decline in star formation.} The ionization falls steadily from the galaxies above the SFMS to the lulling galaxies, which have the lowest [O\,\textsc{iii}]/[O\,\textsc{ii}] ratio (O32) of the sample (median 0.10\,dex, against 0.78\,dex above the SFMS), also at fixed stellar mass, after correction for dust, and not because of metal loss in winds, which would raise it (Sect.~\ref{sec:ion_sequence}). O32 follows the recent change in star formation ($\rho=+0.87$ with $\Delta\log\mathrm{SFR}$) more closely than where a galaxy sits relative to the SFMS ($\rho=+0.69$), which adds no significant information at fixed $\Delta\log\mathrm{SFR}$. [Ne\,\textsc{iii}]/[O\,\textsc{ii}] behaves in the same way ($\rho=+0.80$), and nearly-MS galaxies whose star formation has recently risen are as highly ionized as those above the SFMS (Sect.~\ref{sec:ion_dlogsfr}).

\item \emph{The ionized gas of lulling galaxies differs from that of typical high-redshift galaxies.} Unlike O32, [O\,\textsc{iii}]/H$\beta$ follows the decline only weakly: nearly-MS galaxies with declines similar to those of the lulling galaxies retain 0.56--0.84\,dex, whereas the lulling galaxies drop to 0.23--0.34\,dex (Sect.~\ref{sec:ion_dlogsfr}). In the diagram of O32 against R23, the ratio of [O\,\textsc{ii}] and [O\,\textsc{iii}] to H$\beta$, the lulling galaxies lie between the high-redshift population and local star-forming galaxies, a region essentially empty in line-selected samples: only 2 of 182 JADES galaxies at $4<z<8$ have comparably low O32 (Sect.~\ref{sec:r23o32}).

\item \emph{The ionization reflects both the radiation sources and the state of the gas.} O32 depends on both the supply and hardness of ionizing photons, which fall within ${\sim}10$\,Myr of a decline as the most massive stars die. O32 further depends on the density and filling factor of the ionized gas, which the star formation of the preceding ${\sim}100$\,Myr has lowered by consuming, expelling and dispersing gas. $\Delta\log\mathrm{SFR}$ thus measures the balance between the current photon supply and this earlier depletion of the gas. The PRISM data cannot separate these terms; resolved [S\,\textsc{ii}] densities can test the density term (Sects.~\ref{sec:ion_dlogsfr} and \ref{sec:ion_source}).

\item \emph{The declines are too fast for the gas to have been consumed by star formation.} Under the adopted, extrapolated depletion-time relation (Sect.~\ref{sec:mechanisms}), the e-folding times of the lulling galaxies, at most 32--66\,Myr, are two to five times shorter than passive exhaustion of the gas allows, even with main-sequence depletion times and no inflow, which at these redshifts is itself hard to achieve. The declines thus favor outflows with mass-loading factors $\eta\approx1$--23, comparable to or up to a few times above those of simulations with explicit stellar feedback, or gas that is kept from forming stars. Since the galaxies are isolated (OASIS Paper~I), interactions are unlikely to dominate (Sect.~\ref{sec:mechanisms}). This points to internal feedback, rather than external processes, as the driver of the declines.

\item \emph{Sustaining the suppression is harder than starting it.} In simulations the expelled gas returns quickly, and suppressed phases as long as ours occur mainly at lower masses. Either supernova feedback keeps the gas out for longer than simulations predict, or additional mechanisms are at work, such as radiation pressure or AGN feedback. Our spectra show no clear evidence that an AGN dominates the suppression or is active at the epoch of observation, but some features, such as the H$\alpha$ profiles of the lulling galaxies, are consistent with an AGN contribution (Sect.~\ref{sec:mechanisms}).

\end{itemize}

Together, these results show that the suppression is too rapid for passive exhaustion of the gas supply, lasts as long as or longer than in most simulations, and is accompanied by a low ionization state that tracks the decline in star formation itself and, as OASIS Paper~I found, by little dust. This favors feedback acting on the gas already in the galaxies, either by expelling it (ejective feedback) or by keeping it from forming stars, for instance by heating it or by inducing turbulence. If their low dust content reflects a low gas content, the latter is disfavored, pointing to ejective feedback.

More broadly, our findings change how the quiet phases of bursty star formation should be viewed. They are not brief flickers between bursts but episodes lasting of order 100\,Myr, from which galaxies recover faster than they enter them, and which a typical galaxy at these masses passes through once or twice while growing thirtyfold. Because simulations produce such long phases mainly at lower masses, their duration, together with the speed of the declines into them, offers a sharp test of feedback models at the masses where JWST now finds them. The gas follows the same cycle: since its ionization tracks the recent change in star formation, the extreme ionization reported for line-selected galaxies at $z>4$ characterizes galaxies on and above the SFMS rather than the population across the full burst cycle. Suppressed galaxies, which such samples miss, are therefore an essential laboratory for studying the ISM across all phases of the cycle. Larger spectroscopic samples with a well-characterized selection function can turn our order-of-magnitude duty cycle into a measurement and, together with the timescales of decline and suppression measured here, constrain the power spectrum of star-formation variability at cosmic dawn \citep{caplar+tacchella2019,Wan+2024}.

Two sets of observations can test the feedback picture directly. Deep [C\,\textsc{ii}] observations of stacked or strongly lensed systems can constrain the gas content itself (Sect.~\ref{sec:mechanisms}). Medium-resolution grating spectroscopy, which our Cycle~5 program LULLs (GO~10149) will provide, can measure the gas density and outflows directly, help distinguish stellar- from AGN-driven ionization, and test for Type 1 AGN. The quiet phases of the burst cycle, long missed because their emission lines are faint, may hold the most direct record of how feedback regulates the growth of the first galaxies.

\begin{acknowledgements}
TJL gratefully acknowledges support from the Swiss National Science Foundation through a SNSF Mobility Fellowship and from the NASA/JWST Program OASIS (PID \#5997). FDE acknowledges support by the Science and Technology Facilities Council (STFC), by the ERC through Advanced Grant 695671 ``QUENCH'', and by the UKRI Frontier Research grant RISEandFALL.
WMB gratefully acknowledges support from DARK via the DARK fellowship. This work was supported by a research grant (VIL54489) from VILLUM FONDEN.
SA acknowledges grants PID2021-127718NB-I00 and PID2025-169195NB-I00 funded by the Spanish Ministry of Science and Innovation/State Agency of Research (MICIN/AEI/10.13039/501100011033).
AJB and JC acknowledge funding from the ``FirstGalaxies'' Advanced Grant from the European Research Council (ERC) under the European Union's Horizon 2020 research and innovation programme (Grant agreement No.\ 789056).
SC acknowledges support by the European Union's HE ERC Starting Grant No.\ 101040227 -- WINGS.
QD acknowledges support from a PhD studentship awarded by Trinity College, University of Cambridge.
JMH acknowledges support from the Evolving Universe Fellowship, which is made possible by a generous donation from Dr.\ Keiko Miwa Ross; JMH also acknowledges support from JWST Program \#8544.
YI is supported by JSPS KAKENHI Grant No.\ 24KJ0202.
BER and BDJ acknowledge support from the JWST/NIRCam Science Team contract to the University of Arizona, NAS5-02105, and JWST Program 3215.
PR acknowledges support from the University of Texas at Austin Cosmic Frontier Center.
ST acknowledges support by the Royal Society Research Grant G125142.
JAAT acknowledges support from the Simons Foundation and JWST programs 2959 and 3215. Support for programs 2959 and 3215 was provided by NASA through a grant from the Space Telescope Science Institute, which is operated by the Association of Universities for Research in Astronomy, Inc., under NASA contract NAS 5-03127.
H\"U acknowledges support by the Max Planck Society through the Lise Meitner Excellence Program. H\"U acknowledges funding by the European Union (ERC APEX, 101164796). Views and opinions expressed are however those of the authors only and do not necessarily reflect those of the European Union or the European Research Council Executive Agency. Neither the European Union nor the granting authority can be held responsible for them.
JW gratefully acknowledges support from the Cosmic Dawn Center through the DAWN Fellowship. The Cosmic Dawn Center (DAWN) is funded by the Danish National Research Foundation under grant No.\ 140.
The authors acknowledge use of the lux supercomputer at UC Santa Cruz, funded by NSF MRI grant AST 1828315.
This work is based on observations made with the NASA/ESA/CSA \textit{James Webb Space Telescope}. The data were obtained from the Mikulski Archive for Space Telescopes at the Space Telescope Science Institute, which is operated by the Association of Universities for Research in Astronomy, Inc., under NASA contract NAS 5-03127 for \jwst. These observations are associated with program GO~5997. Target selection used JADES NIRCam imaging.
\end{acknowledgements}

\bibliographystyle{aa}
\bibliography{Bib_TL}

\begin{appendix}

\section{Excitation diagnostics}\label{app:bpt}

\begin{figure}
    \centering
    \includegraphics[width=\linewidth]{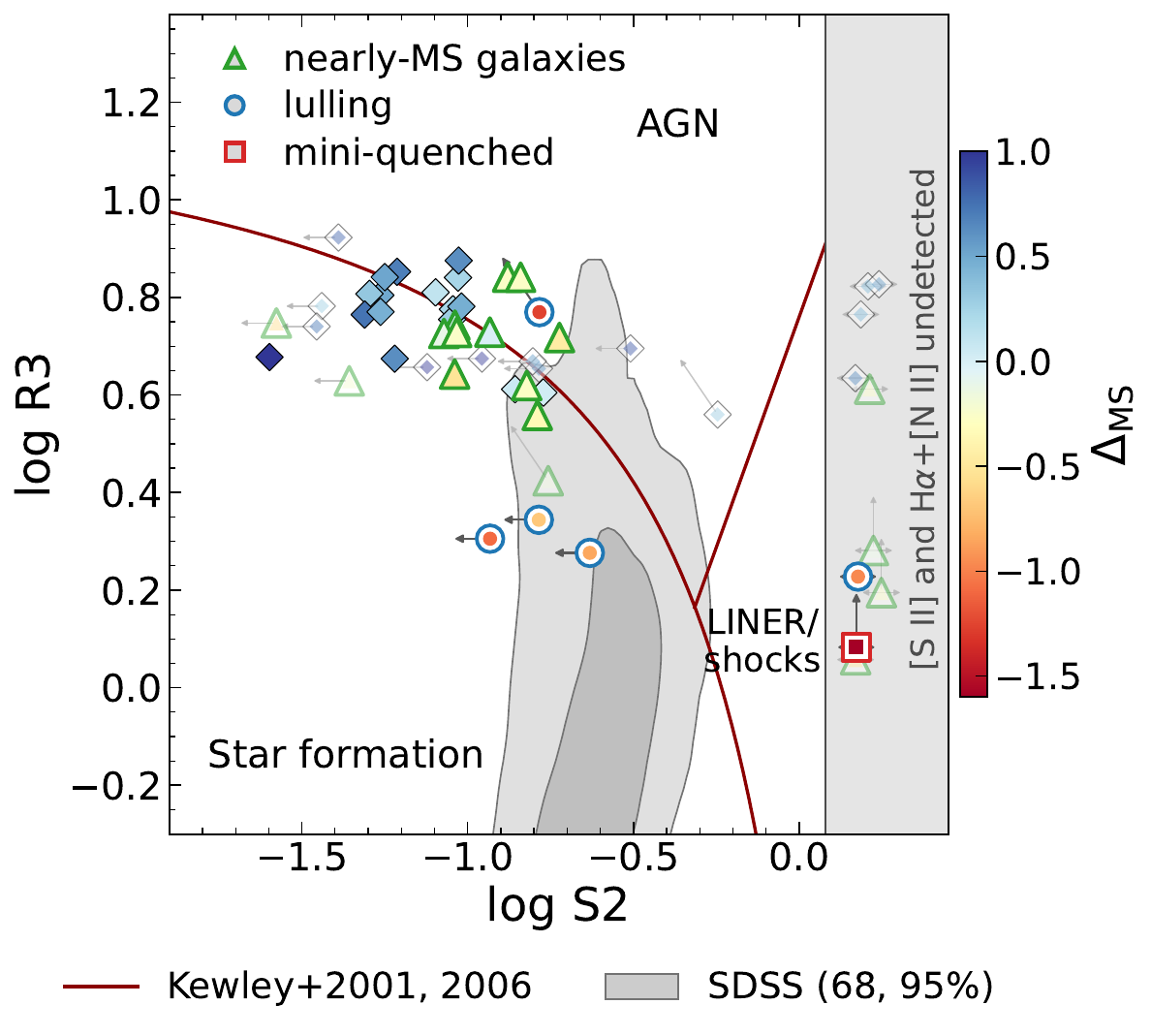}
    \caption{
    \textbf{S2 excitation diagram for the $z>4$ OASIS galaxies.}
    R3 (Eq.~\eqref{eq:R3}) against
    $\mathrm{S2}=[\mathrm{S\,II}]\lambda\lambda6716,6731/(\mathrm{H}\alpha+[\mathrm{N\,II}])$, with the
    maximum-starburst line of \citet{Kewley2001} and the Seyfert--LINER division of
    \citet{Kewley2006}. Both [S\,\textsc{ii}] lines and H$\alpha$+[N\,\textsc{ii}] are blended at
    the PRISM resolution, and the demarcations, defined for H$\alpha$ alone, are applied to the
    blend. Gray contours enclose 68 and 95 per cent of
    the SDSS DR7 emission-line galaxies. Fill color gives
    $\Delta_{\mathrm{MS}}$; marker shape and edge color denote the class as in
    Fig.~\ref{fig:SFR-mass}. Open symbols with arrows are limits; they are faded except for the
    lulling and mini-quenched galaxies. Galaxies with neither [S\,\textsc{ii}] nor H$\alpha$+[N\,\textsc{ii}] detected are
    placed, with double-headed arrows, in the gray band beyond the solid line at the edge of the
    axis; their position within the band carries no information.
    Galaxies uninformative on both axes are omitted. The
    [N\,\textsc{ii}] and [O\,\textsc{i}] BPT diagrams are not shown: at the PRISM resolution [N\,\textsc{ii}]
    cannot be separated from H$\alpha$, and [O\,\textsc{i}] is detected in only eight galaxies,
    none of them suppressed.
    }
    \label{fig:BPT_app}
\end{figure}

\begin{figure}
    \centering
    \includegraphics[width=\linewidth]{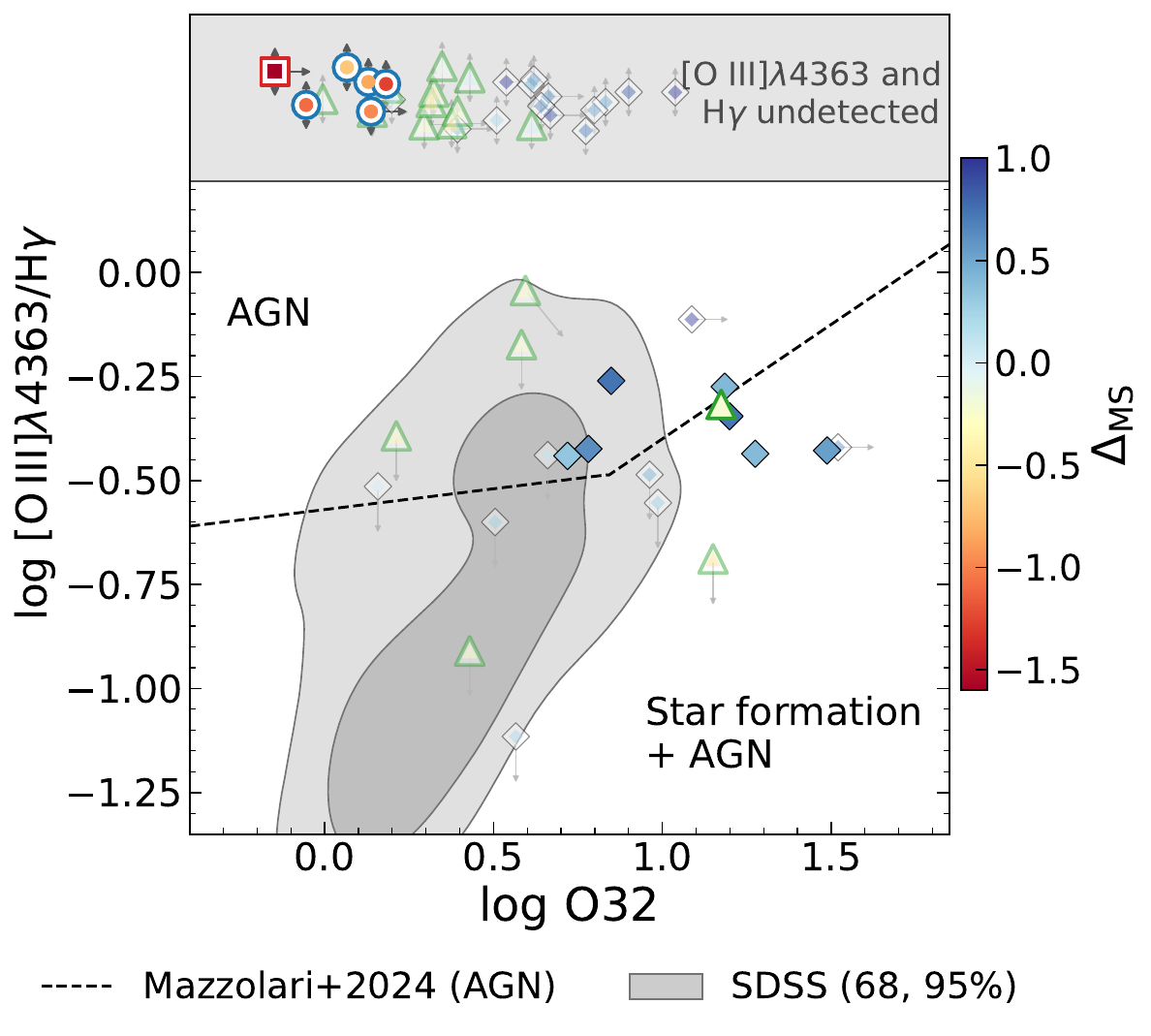}
    \caption{
    \textbf{Auroral-line diagnostic for the $z>4$ OASIS galaxies.}
    The ratio [O\,\textsc{iii}]$\lambda4363$/H$\gamma$ against
    O32 (Eq.~\eqref{eq:O32}), with the AGN
    demarcation of \citet{Mazzolari+2024} (dashed) and local SDSS galaxies (gray contours).
    This demarcation is one-sided: sources above it are AGN-dominated with high confidence,
    while those below may be star-forming or AGN. Galaxies with neither [O\,\textsc{iii}]$\lambda4363$
    nor H$\gamma$ detected are placed in the gray band at the top; their position within the band carries
    no information. Other symbols as in Fig.~\ref{fig:BPT_app}.
    }
    \label{fig:BPT_app_o3hg}
\end{figure}

Figures~\ref{fig:BPT_app} and \ref{fig:BPT_app_o3hg} show the two excitation diagnostics that remain informative at the
PRISM resolution, with
\begin{equation}
  \mathrm{S2} = \frac{[\mathrm{S\,II}]\lambda\lambda6716,6731}{\mathrm{H}\alpha+[\mathrm{N\,II}]\lambda\lambda6548,6584},
  \label{eq:S2}
\end{equation}
where both the numerator and the denominator are blends at the PRISM resolution. In the \sii\ diagram, three of the lulling galaxies (108487, 309077 and
120014) lie in the star-forming region: their \sii\ is undetected, and moving them to
lower \sii/(H$\alpha$+\nii) only increases their distance from the demarcation. For 202438 the
undetected \hb\ gives only a lower limit on \oiii/\hb, which places it close to the
maximum-starburst line, and for 114769, at $z=7.5$, \sii\ and H$\alpha$ lie beyond the PRISM wavelength range. The
nearly-MS galaxies lie at higher \sii/(H$\alpha$+\nii) than those above the SFMS, by
$0.24$\,dex in the median ($-0.91$ against $-1.15$; Gehan $p=0.017$). Of the 26 galaxies with
\sii\ detected, 11 lie above the \citet{Kewley2001} line, four of the ten nearly-MS galaxies
and seven of the sixteen above the SFMS, most of them within $\sim0.1$\,dex of it. The blending
of H$\alpha$ with \nii\ shifts all galaxies to lower ratios, i.e., toward the star-forming
region, by $\log(1+[\mathrm{N\,II}]/\mathrm{H}\alpha)\lesssim0.1$\,dex at the low metallicities
expected here; the three lulling galaxies above remain in the star-forming region even after
such a shift.

The [O\,\textsc{iii}]$\lambda4363$ diagram (Fig.~\ref{fig:BPT_app_o3hg}) is limited by the auroral line, which is detected in
ten galaxies, none of them lulling or mini-quenched. Five of these lie above the
\citet{Mazzolari+2024} demarcation, all above the SFMS; for four
of them the margin is $\leq0.07$\,dex or O32 is only a lower limit. For the lulling galaxies,
whose [O\,\textsc{iii}]$\lambda4363$ and H$\gamma$ are both undetected, the diagram constrains
only O32, which places them at the low-ionization end of the sample, as in
Fig.~\ref{fig:ion_dms}.

As discussed in Sect.~\ref{sec:ion_source}, the local demarcations do not transfer directly to $z\gtrsim4$, so neither diagram can confirm or exclude a Type~2 AGN or shock contribution in the nearly-MS systems, several of which lie close to the demarcations. We expect this degeneracy to be weakest at low ionization parameter, where the lulling galaxies lie (Fig.~\ref{fig:ion_dms}); for them the \sii\ diagram favors star formation in three of the five, although their low metallicities still limit the diagnostic, while for the other two the data are inconclusive. Resolving these cases requires grating spectroscopy.

\section{Spectra of the rising nearly-MS galaxies}\label{app:rising}

\begin{figure*}
  \centering
  \begin{subfigure}[t]{0.48\textwidth}
    \centering
    \includegraphics[width=\linewidth]{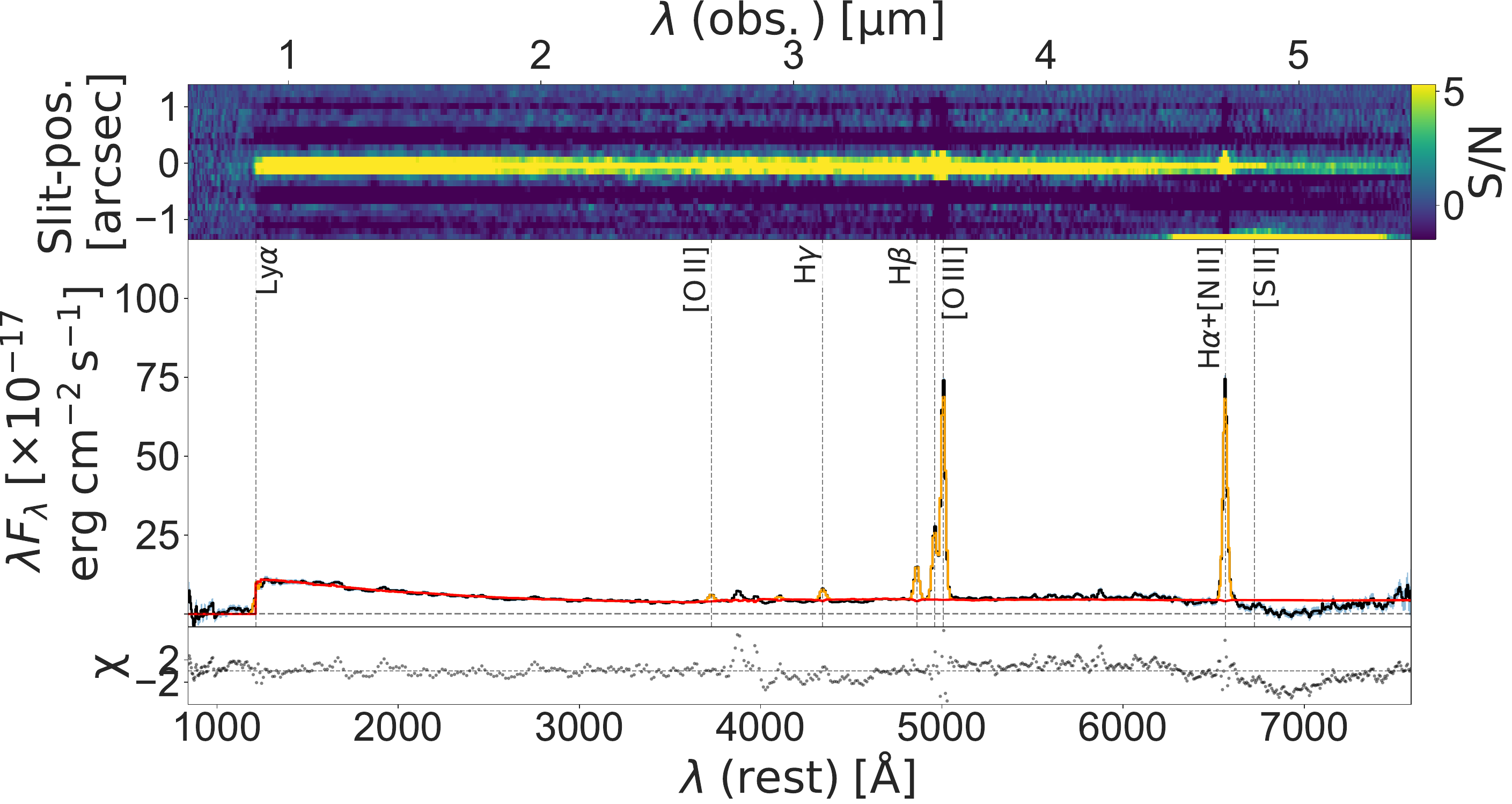}
    \caption{632746 at $z=6.17$; $\log_{10}(M_\star/M_\odot)=9.7$, $\Delta\log\mathrm{SFR}=+0.45$.}
  \end{subfigure}\hfill
  \begin{subfigure}[t]{0.48\textwidth}
    \centering
    \includegraphics[width=\linewidth]{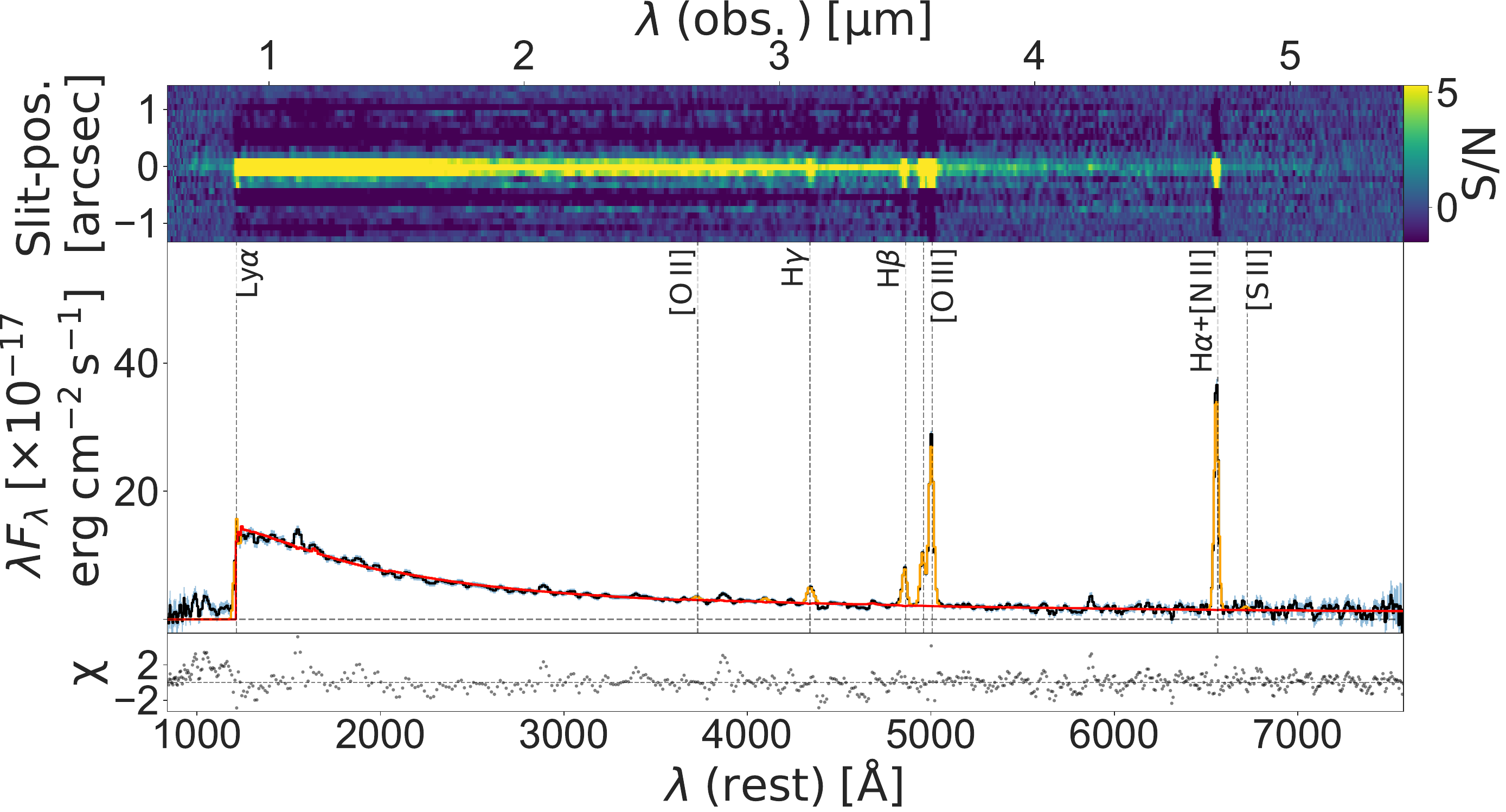}
    \caption{214959 at $z=6.18$; $\log_{10}(M_\star/M_\odot)=9.0$, $\Delta\log\mathrm{SFR}=+0.28$.}
  \end{subfigure}
  \caption{\textbf{Two nearly-MS galaxies caught in a rising phase.} Top: signal-to-noise ratio of the two-dimensional NIRSpec/PRISM spectrum. Middle: one-dimensional spectrum (black) and its noise (blue shading), with the best-fit \ppxf\ model (red) and the fitted emission lines (yellow); dashed lines mark the rest-frame wavelengths of the main nebular lines. Bottom: residuals normalized by the noise. Both show strong [O\,\textsc{iii}] and H$\alpha$ emission on a blue continuum, with light-weighted ages of 22 and 7\,Myr and Balmer breaks of $1.32\pm0.06$ and $1.01\pm0.05$. The stellar continuum independently shows the rise (Sect.~\ref{sec:ion_dlogsfr}); an older population may be outshone by the recent upturn in star formation.}
  \label{fig:rising}
\end{figure*}

Figure~\ref{fig:rising} shows the spectra of the two rising nearly-MS galaxies discussed in Sects.~\ref{sec:tracks} and \ref{sec:ion_dlogsfr}.

\section{Derivation of the elapsed times}
\label{app:tq}

\begin{figure*}
    \centering
    \includegraphics[width=0.9\linewidth]{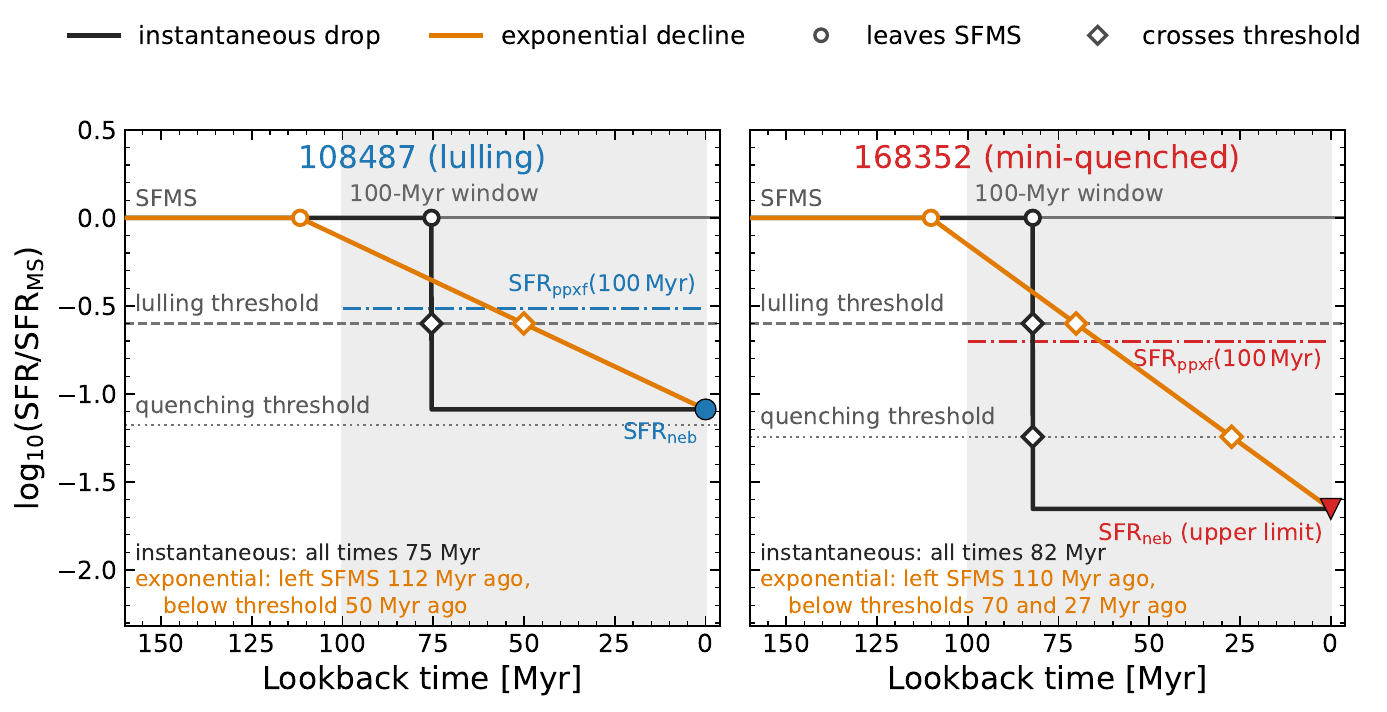}
    \caption{
    \textbf{Instantaneous and exponential declines for a lulling and a mini-quenched galaxy.}
    Star-formation rate relative to the SFMS against lookback time for 108487 (left) and 168352 (right; its nebular rate is an upper limit). Both declines reproduce the present nebular rate and the 100-Myr average from \ppxf\ (dash-dotted); the gray band marks the averaging window, and the dashed and dotted lines the lulling and quenching thresholds. Circles mark where each decline leaves the SFMS and diamonds where it crosses a threshold (Eq.~\eqref{eq:tx}). For an instantaneous drop, all three times coincide; an exponential decline leaves the SFMS earlier but crosses each threshold more recently.
    }
    \label{fig:tracks}
\end{figure*}

Section~\ref{sec:tracks} combines the two measured star-formation rates into
an estimate of how long each suppressed galaxy has been off the main sequence and below the lulling and quenching thresholds.
This appendix sets out the derivation.

\paragraph{Instantaneous drop.} Let a galaxy form stars at the main-sequence rate
$\mathrm{SFR_{MS}}$ up to a lookback time $t_q$, and at the constant reduced
rate $\mathrm{SFR_{neb}}$ from then until the present,
\begin{equation}
    \mathrm{SFR}(t) =
    \begin{cases}
        \mathrm{SFR_{MS}},  & t_q < t \le T, \\[2pt]
        \mathrm{SFR_{neb}}, & 0 \le t \le t_q,
    \end{cases}
\end{equation}
with $t$ a lookback time and $T = 100$\,Myr the \ppxf\ averaging window.

\paragraph{Mass balance.} The mass formed within the window is the sum over
the two intervals,
\begin{equation}
    M(<T) = \mathrm{SFR_{MS}}\,(T - t_q) + \mathrm{SFR_{neb}}\,t_q ,
\end{equation}
and the quantity \ppxf\ reports is the mean rate across it,
\begin{equation}
    \mathrm{SFR_{ppxf}}(T) \equiv \frac{M(<T)}{T}
    = \frac{\mathrm{SFR_{MS}}\,(T - t_q) + \mathrm{SFR_{neb}}\,t_q}{T} .
    \label{eq:mean}
\end{equation}

\paragraph{Normalization.} Dividing Eq.~\eqref{eq:mean} by
$\mathrm{SFR_{MS}}$ and expressing both rates in units of the sequence,
\begin{equation}
    a \equiv \frac{\mathrm{SFR_{ppxf}}(T)}{\mathrm{SFR_{MS}}}, \qquad
    n \equiv \frac{\mathrm{SFR_{neb}}}{\mathrm{SFR_{MS}}}
      = 10^{\Delta_{\rm MS}} ,
\end{equation}
where $\mathrm{SFR_{ppxf}}(T)$ is the median over the \ppxf\ bootstrap
realizations and $\mathrm{SFR_{MS}}$ is evaluated at the galaxy's present
mass and redshift, gives
\begin{equation}
    a\,T = (T - t_q) + n\,t_q = T - t_q\,(1 - n) .
\end{equation}

\paragraph{Solution.} Collecting $t_q$,
\begin{equation}
    t_q = T\,\frac{1-a}{1-n} .
    \label{eq:tq}
\end{equation}
Equation~\eqref{eq:tq} applies for $n \le a \le 1$, which places $t_q$ in
$[0,T]$. The limits behave as required: $a=1$ gives $t_q=0$ for a galaxy
whose 100\,Myr average is still on the sequence, and $a=n$ gives $t_q=T$ for
one suppressed throughout the window. Outside this range the model does not
apply. For $a>1$ the galaxy lay \emph{above} the sequence on average over
the window, so its decline, if any, began from above it; for $a<n$ the
100\,Myr average lies below the present rate, which indicates either a
recent rise or suppression that predates the window altogether. The case
$n \to 1$ is degenerate, since a galaxy still on the sequence has no
departure to date.

\paragraph{Exponential decline.} For a decline that is linear in $\log\mathrm{SFR}$, $\mathrm{SFR}(t)=\mathrm{SFR_{MS}}\,n^{1-t/t_q}$ for $t\le t_q$, and the mean over the window becomes
\begin{equation}
    a = \begin{cases}
        1-\dfrac{t_q}{T}\left(1-\dfrac{1-n}{\ln(1/n)}\right), & t_q\le T, \\[8pt]
        \dfrac{t_q}{T}\,\dfrac{n^{1-T/t_q}-n}{\ln(1/n)}, & t_q>T,
    \end{cases}
\end{equation}
which we solve numerically for $t_q$. The decline then crosses a threshold $\Delta_{\rm th}$ at the lookback time of Eq.~\eqref{eq:tx}, whereas an instantaneous drop falls below every threshold it crosses at $t_q$. Figure~\ref{fig:tracks} shows both declines for one lulling and one mini-quenched galaxy.

\paragraph{From elapsed times to episode durations.} Let suppressed episodes have
durations distributed as $p(d)$, with mean $\langle d\rangle$ and standard
deviation $\sigma_d$. A galaxy observed at a random moment is caught in an
episode of duration $d$ with probability proportional to $d\,p(d)$, and the
time already elapsed in that episode is uniformly distributed between 0 and
$d$, with mean $d/2$. The mean elapsed time is therefore
\begin{equation}
    \langle t_\mathrm{el}\rangle
    = \frac{\int_0^\infty (d'/2)\,d'\,p(d')\,\mathrm{d}d'}{\int_0^\infty d'\,p(d')\,\mathrm{d}d'}
    = \frac{\langle d^2\rangle}{2\langle d\rangle}
    = \frac{1+c^2}{2}\,\langle d\rangle ,
    \label{eq:tel}
\end{equation}
where $c=\sigma_d/\langle d\rangle$ is the coefficient of variation of the
durations. Since $c^2\geq0$, $\langle d\rangle\leq2\langle t_\mathrm{el}\rangle$
for any $p(d)$, with equality ($c=0$) if all episodes have the same length.
An exponential distribution has $c=1$ and gives $\langle d\rangle=\langle
t_\mathrm{el}\rangle$; distributions no broader than exponential ($c\leq1$)
give $\langle d\rangle\geq\langle t_\mathrm{el}\rangle$. Broader, heavy-tailed
distributions would imply shorter mean durations and hence more episodes
(Eq.~\eqref{eq:duty}).

\end{appendix}
\end{document}